\documentclass[10pt,sigconf,letterpaper]{acmart}

\usepackage{booktabs} %
\usepackage{graphicx}   %
\usepackage{mathtools}
\usepackage{amsmath}
\usepackage{color, colortbl}
\usepackage[nolist]{acronym}
\usepackage{multirow}
\usepackage{euscript}
\usepackage{adjustbox}
\usepackage{array}
\usepackage{appendix}
\usepackage[tableposition=bottom]{caption}
\usepackage{subfig}
\usepackage{placeins}
\usepackage[noabbrev]{cleveref}

\usepackage{hyperref}
\def\Snospace~{\S{}}

\hypersetup{
    colorlinks,%
    citecolor=[rgb]{0.7,0,0},
    filecolor=[rgb]{0.2,0.2,0.2},
    linkcolor=[rgb]{0,0,0.7},
    urlcolor=[rgb]{0.7,0,0},
}

\definecolor{Gray}{gray}{0.9}
\newcommand{\Vs}[1]{\text{\textit{#1}}}  %

\newcolumntype{C}{>{\centering\arraybackslash}m{4.5em}}
\newcolumntype{D}{>{\centering\arraybackslash}m{3.5em}}
\newcolumntype{M}[1]{>{\centering\arraybackslash}m{#1}}
\newcolumntype{P}[1]{>{\centering\arraybackslash}p{#1}}
\newcolumntype{N}{@{}m{0pt}@{}}

\setcopyright{none}

\renewcommand\footnotetextcopyrightpermission[1]{} %

  \acmDOI{}
  \acmISBN{}
  \acmConference{Technical Report}{Mar 2023}{Marina del Rey, CA, USA}
  \acmYear{}
  \copyrightyear{2023}
  \acmArticle{}
  \acmPrice{}

\makeatletter
  \newcommand\EatSpacesHack{\@bsphack\@esphack}
\makeatother

  \renewcommand\comment[1]{\EatSpacesHack}
  \newcommand\yp[1]{\EatSpacesHack}
  \newcommand\gb[2]{\EatSpacesHack}
  \newcommand\reviewfix[1]{\EatSpacesHack}

  \newcommand\PostSubmission[1]{\EatSpacesHack}

  \makeatletter
    \renewcommand\section{\@startsection{section}{1}{\z@}%
      {-.25\baselineskip \@plus -2\p@ \@minus -.2\p@}%
      {.1\baselineskip}%
      {\ACM@NRadjust\@secfont}}
    \renewcommand\subsection{\@startsection{subsection}{2}{\z@}%
      {-.15\baselineskip \@plus -2\p@ \@minus -.2\p@}%
      {.05\baselineskip}%
      {\ACM@NRadjust\@subsecfont}}
    \renewcommand\subsubsection{\@startsection{subsubsection}{3}{\z@}%
      {-.25\baselineskip \@plus -2\p@ \@minus -.2\p@}%
      {-2\p@}%
      {\ACM@NRadjust{\@subsubsecfont\@adddotafter}}}
   \makeatother
  \newcommand{\RelaxFloats}{
  	\renewcommand{\topfraction}{0.9}
  	\renewcommand{\floatpagefraction}{0.9}
  	\renewcommand{\textfraction}{0.1}
  }
  \RelaxFloats

\acrodefplural{AS}[ASes]{Autonomous Systems}
\acrodefplural{VP}[VPs]{Vantage Points}
\acrodef{NAT}[NAT]{Network Address Translation}
\acrodef{CG-NAT}[CG-NAT]{Carrier-Grade NAT}
\acrodefplural{CDN}[CDNs]{Content Delivery Networks}

\begin{document}
\title[Detecting Partial Reachability In The Internet Core]{Detecting Partial Reachability In The Internet Core}

\subtitle{Technical Report: arXiv:2107.11439v4 \\ released 2021-07-23, updated 2022-05-24, 2022-03-20, 2023-10-26}

  \setlength{\dblfloatsep}{-5pt}
  \addtolength{\abovecaptionskip}{-4pt}
  \setlength{\dbltextfloatsep}{0pt}
  \setlength{\abovedisplayskip}{1.5pt plus 1pt minus 1pt}
  \setlength{\belowdisplayskip}{1.5pt plus 1pt minus 1pt}

  \settopmatter{authorsperrow=4}
  \author{Guillermo Baltra}
  \affiliation{%
    \department{USC/ISI}
    \city{Marina del Rey}
    \state{California}
    \country{USA}
  }
  \email{baltra@isi.edu}

  \author{Tarang Saluja}
  \affiliation{%
    \department{Swarthmore College}
    \city{Swarthmore}
    \state{Pennsylvania}
    \country{USA}
  }
  \email{tsaluja1@swarthmore.edu}

  \author{Yuri Pradkin}
  \affiliation{%
    \department{ISI}
    \city{Marina del Rey}
    \state{California}
    \country{USA}
  }
  \email{yuri@isi.edu}

  \author{John Heidemann}
  \affiliation{%
    \department{USC/ISI}
    \city{Marina del Rey}
    \state{California}
    \country{USA}
  }
  \email{johnh@isi.edu}

  \renewcommand{\shortauthors}{Baltra et al.}

\begin{abstract}
The Internet core should seamlessly connect
  home and mobile users to servers, CDNs, and the cloud.
Yet seamless, universal connectivity is challenged in several ways.
Political pressure threatens fragmentation due to de-peering;
  architectural changes such as carrier-grade NAT and firewalls can impede connectivity;
  and operational problems and commercial disputes result in
  unreachability for days or even years.
We must recognize that \emph{partial reachability is a fundamental part of the Internet's core}.
To understand reachability
  we define \emph{peninsulas}, persistent, partial connectivity;
  and \emph{islands}, when one or more computers are partitioned from the main Internet.
We develop new algorithms to
  measure the
  number, size, and duration of peninsulas and islands.
These algorithms follow from a conceptual definition of
  the Internet's core defined by connectivity, not special authority.
\emph{Recognizing peninsulas and islands can improve
  existing measurement systems in several ways.}
We show how to improve the sensitivity of DNSmon,
  removing measurement error and persistent problems are $5\times$ to $9.7\times$ larger
  than its operationally important signal.
Partial connectivity clarifies results seen in
  several outage detection systems
  that otherwise appear self-contradictory.
Our analysis also informs policy questions and confirming
  the international nature of the Internet:
  no single country today unilaterally controls the Internet core,
  but countries can choose to leave.
\emph{We validate and evaluate these algorithms with rigorous measurements
  from two complementary measurement systems},
  one observing 5M networks from a few locations,
  and the other
  a few destinations from 10k locations,
  and with external data.
Results show that
  peninsulas (partial connectivity)
  are about as common as Internet outages,
  quantifying this long-observed problem.
Root causes show that
  most peninsula events (45\%) are routing transients,
  but most peninsula-time  (90\%) is from
  a few long-lived events (7\%).
\end{abstract}

\begin{acronym}[AS]
\acro{AS}{Autonomous System}
\acro{RDNS}{Reverse DNS}
\acro{VP}{Vantage Point}
\end{acronym}

\maketitle

\vspace*{-1ex}

\section{Introduction}
	\label{sec:introduction}

\emph{What \emph{is} the Internet's core?}
In 1982, Smallberg's census of the IPv4 Internet showed 83 telnet-reachable hosts
  in \emph{the} Internet, a globally reachable core.
And in 1995,
  the Federal Networking Council defined ``Internet''
  in 1995 as (i) a global address space,
  (ii) supporting TCP/IP and its follow-ons,
  that (iii) provides services~\cite{nitrd},
  with later work considering DNS~\cite{IAB20a} and IPv6.
But today's Internet is dramatically different than 1995:
Users at home and work access the Internet indirectly through
  \ac{NAT}~\cite{Tsuchiya93a}.
Most access is from mobile devices,
  often behind \ac{CG-NAT}~\cite{Richter16c}.
Many public services are operated from the cloud,
  visible through rented or imported IP addresses,
  but backed
  with complex services built on virtual networks (for example~\cite{Greenberg09a}).
Content is replicated in \acp{CDN}.
Access to each is mediated by firewalls.
Today's Internet services are so seamless
  that technical details recede into the background
  and laypeople consider the web, Facebook, or their mobile phone as their ``Internet''.

\textbf{Challenges:}
Yet \emph{universal reachability in Internet core today is often challenged}.
\emph{Political} pressure
  and threats of disconnection suggest national borders may balkanize the core:
  the 2019 ``sovereign Internet'' law in Russia~\cite{BBC19a,RBC21a,Reuters21a},
  and a national ``Internet kill switch''
  has been debated (including the U.S.~\cite{GovTrack20a} and U.K.),
  and employed~\cite{Cowie11a,Coca18a,Griffiths19a,Taye19a} %
These pressures prompted policy discussions about fragmentation~\cite{Drake16a,Drake22a}.
We suggest that \emph{technical methods can help inform policy discussions}
  and show what is at risk for the global Internet from threats such as de-peering.
We will show that no single country
  can unilaterally control the Internet today  (\autoref{sec:other_applications}).
We also show that de-peering can fragment the Internet
  into pieces (\autoref{sec:internet_partition}).

\emph{Architecturally}, twenty-five years of evolution
  have segmented the Internet,
  services gatewayed through proprietary cloud APIs,
  users increasingly relegated to second-class status as clients,
  often behind \ac{CG-NAT},
  firewalls interrupt connectivity,
  and a world straddling a mix of IPv4 and IPv6.
Architecture sometimes follows politics,
  with
  China's Great Firewall managing international communication~\cite{Anonymous12a,Anonymous14a},
  and
  Huawei proposing ``new Internet'' protocols~\cite{huawei2020}.
We suggest that technical methods to detect
  can help us \emph{reason about changes to Internet architecture},
  both to
  understand the implications of partial address reachability
  and evaluate the maturity of IPv6.

\emph{Operationally}, ISP peering is mature,
  but today
  peering disputes cause long-term  partial unreachability~\cite{ipv6peeringdisputes}.
This unevenness has been recognized and detected experimentally~\cite{Dhamdhere18a},
  and in systems that detect and bypass partial
  reachability~\cite{andersen2001resilient,katz2008studying,katz2012lifeguard}.
We show several operational uses of our work.
We show that \emph{accounting partial reachability can make existing measurement
  systems more sensitive},
  applying these results to widely used RIPE DNSmon (\autoref{sec:dnsmon}).
DNSmon sees
   persistent high query loss (5--8\%)
   in DNS Root Server System~\cite{RootServers16a},
   but most of this loss is due to measurement error or
   persistent partial connectivity, factors that are $5\times$ and $9.7\times$
   larger than the operationally important signal in IPv4 and IPv6.
Our analysis also helps uncertainty in multiple,
  independently developed outage detection systems (\autoref{sec:local_outage_eval}).
All existing outage detection systems encounter
  ``corner cases''~\cite{Schulman11a,quan2013trinocular,Shah17a,richter2018advancing,guillot2019internet}
    and conflicting observations.
We show those are due to partial reachability,
   and we show partial reachability is as common as complete outages  (\autoref{sec:peninsula_frequency}).
Our work also helps quantify the applicability of
  systems that, since 2001, route around partial reachability~\cite{andersen2001resilient,katz2008studying,katz2012lifeguard},
  and show that clouds can improve reliability with egress selection (for example,~\cite{Schlinker17a}).

\textbf{Contributions:}
The first contribution of this paper is to
  \emph{recognize partial reachability is a fundamental part of the Internet's core}.
We define
  \emph{peninsulas}, when a network sees persistent, partial connectivity
  to part of the Internet,
and
  \emph{islands},
  when one or more computers are partitioned from the main Internet.
We develop algorithms to measure each (\autoref{sec:design}).
\emph{Taitao} detects peninsulas
  that often result from peering disputes or long-term firewalls.
Our second algorithm, \emph{Chiloe}, detects islands.
These algorithms are operational,
  able to estimate the presence of peninsulas and islands in
  existing measurement data from two different Internet-wide measurement systems.

A rigorous definition of peninsulas and islands requires
  that we identify \emph{the Internet's core}.
\emph{The Internet's core is the connected component of more than
  50\% of active, public IP addresses that can initiate communication with each other}
  (\autoref{sec:definition}).
This definition has several unique characteristics.
First, requiring bidirectional initiation
  captures the uniform, \emph{peer-to-peer nature
  of the Internet's core}
  necessary for first-class services.
Second, it defines \emph{one, unique} Internet core
  by requiring reachability of more than 50\%---there can be only one
  since multiple majorities
  are impossible.
Finally, unlike prior work, this \emph{conceptual} definition
  avoids dependence on any specific measurement system,
  nor does it depend on historical precedent, special locations, or central authorities.
Although our operational measurements of peninsulas and islands may
  reflect observation error,
  the conceptual Internet core
  defines an asymptote against which our current and future measurements can compare,
  unlike prior definitions from specific systems~\cite{andersen2001resilient,katz2008studying,katz2012lifeguard}.

Our second contribution is to \emph{use peninsulas and islands
  to address current operational questions}.
As described earlier,
  we bring technical light to policy choices
  around national networks (\autoref{sec:other_applications}) and de-peering (\autoref{sec:internet_partition}).
We improve sensitivity of RIPE Atlas' DNSmon~\cite{Amin15a} (\autoref{sec:dnsmon}),
  resolve corner cases in outage detection (\autoref{sec:peninsula_frequency}),
  and quantify opportunities for route detouring.

Our final contribution is
  to \emph{support these claims with rigorous measurements}
  from two measurement systems.
We evaluate our new algorithms
  with publicly available, existing measurements
  of connectivity to 5M networks
    from six \acp{VP} over multiple years~\cite{quan2013trinocular}.
While a handful of locations cannot represent the entire Internet,
  each observer scans most of the ping-responsive Internet
  from a unique geographic and network location,
  providing a wide range of results over time.
Our analysis shows that combinations of any three independent \acp{VP}
  provide a result that is statistically indistinguishable from the asymptote
  \autoref{sec:peninsula_frequency}.
We show our algorithms
  provide consistent results,
  offering reproducible and useful estimates
  of Internet reachability and partial connectivity.
We also validate
  interesting events
  with selective traceroutes.

We also evaluate about 10k globally distributed \acp{VP} (RIPE Atlas,~\cite{Ripe15c})
  observing connectivity to 13 anycast destinations (the Root Server System~\cite{RootServers16a}).
These observations from thousands of locations over multiple years
  validate the occurrence of rare events like islands,
  and demonstrate how pervasive peninsulas are.
They confirm our results of Internet-wide scans,
  and allow us to tune DNSmon, as described earlier.

\textbf{Artifacts:}
All of the data used  (\autoref{sec:data_sources})
  and
    created~\cite{ANT22b}
  in this paper
  is available at no cost.
We review ethics in detail in  \autoref{sec:research_ethics},
  but our bulk analysis of IP addresses does not associate them with individuals.
Our work was IRB reviewed and identified as non-human subjects research
  (USC IRB IIR00001648).

This technical report was first released in July 2021.
In May 2022 it was updated with several additions:
a more careful definitions in \autoref{sec:definition} and \autoref{sec:internet_landscape},
new information about island durations \autoref{sec:islands_duration}
  and sizes \autoref{sec:islands_sizes},
expanded applications in \autoref{sec:applications}
  and \autoref{sec:other_applications} and \autoref{sec:internet_partition},
considerable additional details and supporting data in appendices, and
many writing improvements.
In October 2023 we updated it to address typos and missing references.

\section{The Problem of Partial Reachability}
	\label{sec:problem}

To understand partial reachability, we first must define \emph{what}
  is being reached.
Our goal is to rigorously define \emph{the Internet's core}.
Historically it was assumed that ``the Internet'' was whatever was reached
  by the global IPv4 address space over a common TCP/IP protocol~\cite{Cerf74a,Postel80b,nitrd}.
However,
  today's challenges impose two new requirements.
First, a definition should be both
  \emph{conceptual} and \emph{operational}~\cite{scientific_methods}.
Our conceptual definition in \autoref{sec:definition}
  articulates \emph{what} we would like to observe.
In \autoref{sec:design} we operationalize it,
  describing
  \emph{how} actual measurement systems can use it
  to evaluate peninsulas and islands relative to the Internet's core.
The conceptual definition
  suggests a limit that implementations can approach
  (\autoref{sec:peninsula_frequency}),
  even if it cannot be directly implemented.
Prior definitions are too vague to operationalize.

Second, a definition must give both sufficient \emph{and}
  necessary conditions to be part of the Internet's core.
Prior work gave
  properties the core must have (sufficient conditions, like supporting TCP).
We add necessary conditions that indicate
  when networks \emph{leave} the Internet's core.

Finally, the ``challenges'' listed in \autoref{sec:introduction}
  motivate \emph{why} we define the Internet's core:
  doing so will help answer political, architectural, and operational questions.

\subsection{The Internet: A Conceptual Definition}
	\label{sec:definition}

We define the Internet core as \emph{the connected component
  of more than 50\% of active, public IP addresses that can initiate communication with each other}.
Computers behind NAT and in the cloud are on \emph{branches},
  participating but not part of the core, typically with dynamically allocated
  or leased public IP addresses.
This conceptual definition gives \emph{two} Internet cores,
  one for the IPv4 address space and one for IPv6.

This definition follows from the terms
  ``interconnected networks'', ``IP protocol'', and ``global address space''
  used in prior, informal definitions~\cite{Cerf74a,Postel80b,nitrd}---they all share the common assumption
  that two computers on the Internet should be able to communicate
  directly with each other at the IP layer.

We formalize ``an agreement of networks to interconnect''
  by considering reachability over
  public IP addresses:
  addresses $x$ and $y$ are interconnected if traffic from $x$ can reach $y$
    and vice versa (that is: $x$ and $y$ can reach each other).
Networks are groups of addresses that can reach each other.

\textbf{Why More than 50\%?}
We take as an %
  axiom that there should be \emph{one Internet core},
  or reason a single Internet core no longer exists.
Thus we require a definition to unambiguously identify ``the'' Internet core
  given conflicting claims.

We require that the Internet core include more than 50\% of active addresses
  so that the majority can settle conflicting claims.
Only one group can control a majority of addresses,
  while any smaller fraction could allow two groups to make
  valid claims.
A strict majority ensures that there is always a well-defined Internet core
  even if a major nation (or group of nations) chose to secede,
  because it defines a unique, unambiguous partition that keeps the Internet.

Prior discussions of the Internet identified ``Tier-1'' ISPs,
  but strict definitions of that term remain elusive.
Our definition allows us to reason about differences between what ISPs see,
  even when ISPs have long-term peering disputes and partial connectivity.

This definition suggests that it is possible for the Internet to fragment:
  if the current Internet breaks into three disconnected components
  when none has a majority of active addresses.
Such a result would end a single, global Internet.

\textbf{Why all and active addresses?}
In each of IPv4 and IPv6
  we consider all addresses equally. %
The Internet is global,
  and was intentionally designed without a hierarchy~\cite{clark1988design}.
Our definition should not create a hierarchy
  or designate special addresses by age or importance,
  consistent with trends towards Internet
  decentralization~\cite{dinrg}.

\emph{Active} addresses as blocks that are reachable, as defined below.
Our goal is to exclude the influence of large allocated, but unused, space.
Large unused space is present in IPv4 legacy /8 allocations
  and in large new IPv6 allocations.

\textbf{Reachability with Protocols and Firewalls:}
This conceptual definition allows for different definitions of reachability.
Reachability can be tested through measurements with specific protocols,
  such as ICMP echo-request (pings),
  or TCP or UDP queries.
Such a test will result in an operational realization
  of our conceptual definition.
Particular tests will differ
  in how closely each approaches the conceptual ideal.
In \autoref{sec:peninsula_frequency} we examine how well one test converges.

Our conceptual definition considers reachability,
  observations of this potential are complicated
  by
  firewalls (which are sometimes conditional or unidirectional).
Answers may vary with different protocols or observations at different times.
One could define reachability narrowly,
  or using firewall-friendly protocols.
Measurement allows us to evaluate
  policy-driven unreachability in \autoref{sec:country_peninsulas}.

Our operational data uses ICMP echo requests
  (\autoref{sec:data_sources}),
  following
  prior work that compared alternatives~\cite{Bartlett07d,quan2013trinocular,durumeric2014internet}
  and showed ICMP provides better coverage than alternatives,
  and can avoid attenuation from rate limiting~\cite{Guo18a}.
\textbf{Why reachability and not applications?}
Users care about applications, and a user-centric view
  might emphasize reachability of HTTP or to Facebook
  rather than at the IP layer.
We recognize this attention,
  but intentionally measure reachability at the IP layer
  as a more fundamental concept.
IP has changed only twice since 1969 with IPv4 and IPv6,
  but dominant applications ebb and flow,
  and some applications
  extend beyond the Internet.
(E-mail has been transparently relayed to UUCP and FidoNet,
  and the web to pre-IP mobile devices with WAP.)
Future work may look at applications,
  but we see IP-level reachability as an essential starting point.

\textbf{Why bidirectional reachability?}
Most computers today are on branches off the core,
  behind \ac{NAT} or in the cloud.
While such computers are useful as Internet clients,
  they are second-class servers.
NAT'ed computers may use protocols
  such as STUN~\cite{Rosenberg03a} to rendezvous through the core,
  or UPnP~\cite{Miller01a} or PMP~\cite{Cheshire13d} that reconfigure a NAT on the core.
Huge services run in the cloud by leasing public IP addresses from the cloud operator
  or importing their own (BYOIP).

Similarly, services may be implemented as many computers that share a single public
  IP address with load balancing or IP anycast~\cite{Partridge93a},
  perhaps with cloud-based address translation~\cite{Greenberg09a}.
Computers with only application-level availability
  are also not fully part of the Internet core.

\subsection{The Internet Landscape}
\label{sec:internet_landscape}

Our definition of the Internet's core highlights
  its ``rough edges''.
Using our conceptual definition of the Internet as the fully connected component (\autoref{sec:definition}),
  we identify three specific problems:
   an address $a$ is
   a \emph{peninsula} when it has partial connectivity to the Internet's core,
   an \emph{island} when it cannot reach any of the Internet's core,
   and an \emph{outage} only when it is off.

\subsubsection{Outages}
\label{sec:outages}

A number of groups have examined Internet outages~\cite{Schulman11a,quan2013trinocular,richter2018advancing,guillot2019internet}.
These systems observe the IPv4 Internet and identify networks
  that are no longer reachable---they have left the Internet.
Often these systems define outages operationally
  (network $b$ is out because none of our \acp{VP} can reach it).
Conceptually, an outage is when all computers
  in a block are off,
  such as due to a power outage.
When the computers are on but cannot reach the Internet,
  we consider them islands, a special case of outage that we defined next.

\subsubsection{Islands: Isolated Networks}
	\label{sec:island}

An \emph{island} is a group of public IP addresses
  partitioned from the Internet's core,
  but still able to communicate among themselves.
Operationally outages and islands are both unreachable from an external \ac{VP},
  but computers in an island can reach each other.

Islands occur when an organization that has a single connection
  to the Internet loses its router or link to its ISP\@.
A single-office business may become an island
  when its router's upstream connection fails,
  but computers in the office can still reach each other and in-office servers.
In the smallest case, an \emph{address island},
  a computer can ping only itself.
Islands are a special case of outages, and
  we suspect that most outages are actually temporary islands.
{2023-02-13}

\begin{figure}
  \includegraphics[width=0.9\columnwidth]{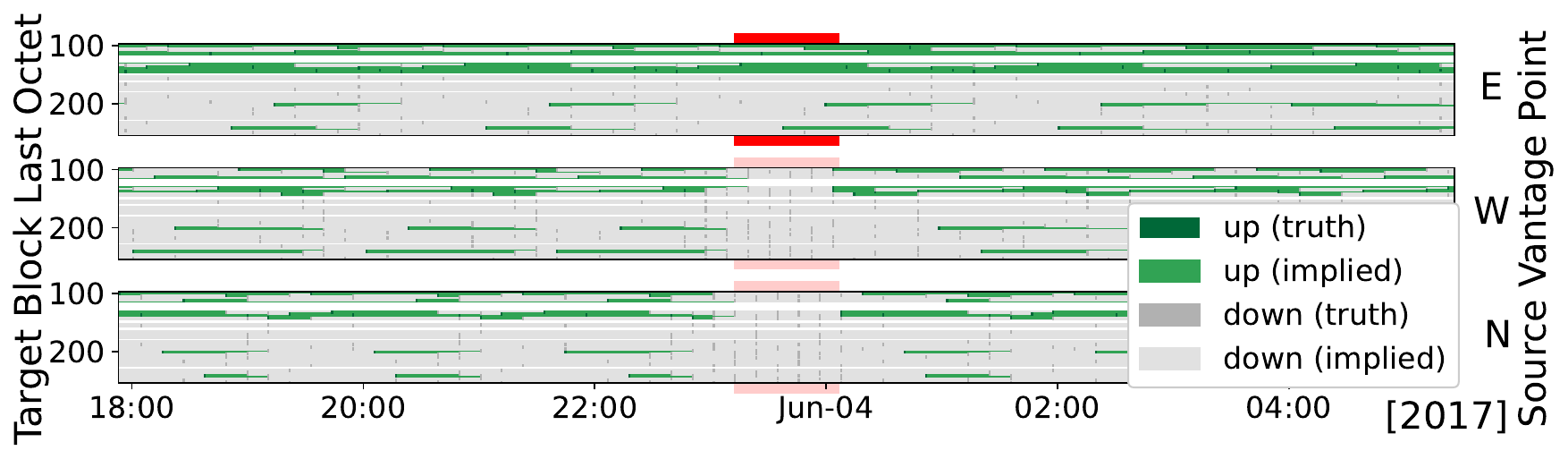}
  \caption{\small A 1-hour island %
     where
             block 65.123.202.0/24
     reaches itself from \ac{VP} E (top)
     but not other VPs (W and N shown).
         (2017q2)
         }
  \label{fig:a28all_raw_417bca00}
  \vspace*{-4ex}
\end{figure}

\textbf{A Brief Island:}
\autoref{fig:a28all_raw_417bca00} shows an example of an island we have observed.
In this graph, each strip shows a different \ac{VP}'s view of
  the last 156 addresses from the same IPv4 /24 block
  over 12 hours.
In each strip, the darkest green dots show positive responses of that address to
  an ICMP echo request (a ``ping'') from that observer,
  and medium gray dots indicate a non-response to a ping.
We show inferred state as lighter green or lighter gray until the next probe.
We show 3 of the 6 \acp{VP}, with probes intervals of about 11 minutes
  (for methodology, see \autoref{sec:data_sources}).

The island starts at
  2017-06-03t23:06Z
  and is indicated by the red bar in the middle of the graph,
  where \ac{VP} E continues to get positive responses from several other addresses
  (the continuous green bars along the top).
By contrast, the other 5 \acp{VP} (2 \acp{VP} here, others in
  \autoref{sec:additional_island})
  show many non-responses
  during this period.
For this whole hour,
  \ac{VP} E and this network are part of an island,
  cut off from the rest of the Internet and the other \acp{VP}.
Although this island is brief and affects only this /24 block
  we have also seen country-sized islands (in \autoref{sec:country_sized_islands} for space).

\subsubsection{Peninsulas: Partial Connectivity}
	\label{sec:peninsula_definition}

Link and power failures create islands, but a more pernicious problem
  is \emph{partial} connectivity,
  when one can reach some destinations,
  but not others.
We call a group of public IP addresses with partial connectivity to the Internet's core
  a \emph{peninsula}.
    (In a geographic peninsula, the mainland may be visible over water, but reachable only with a detour.  In a network peninsula, routing between two points may require a relay through a third location.) %
Peninsulas occur when
  some upstream providers of a multi-homed network
  accept traffic but then drop it due to outages, peering disputes, or firewalls.
 Peninsula existence has long been recognized,
  with overlay networks designed to route around them
  in RON~\cite{andersen2001resilient}, Hubble~\cite{katz2008studying}, and LIFEGUARD~\cite{katz2012lifeguard}.

\textbf{Examples in IPv6:}
An example of a persistent peninsula is
  the IPv6 peering dispute between Hurricane Electric (HE) and Cogent.
These ISPs decline to peer in IPv6,
  nor are they willing to forward their IPv6 traffic through
  another party.
This problem was noted in 2009~\cite{ipv6peeringdisputes} and
  is visible as of June 2020 in DNSMon~\cite{dnsmon} (\autoref{sec:dnsmon}).
We confirm unreachability
  between HE and Cogent users in IPv6 with traceroutes
  from looking glasses~\cite{he_looking_glass,cogent_looking_glass}
  (HE at 2001:470:20::2 and Cogent at 2001:550:1:a::d).
Neither can reach their neighbor's server,
  but both reach their own.
(Their IPv4 reachability is fine.)

Other IPv6 disputes are Cogent with Google~\cite{google_cogent}, and
 Cloudflare with Hurricane Electric~\cite{cloudfare_he}.
Disputes are often due to an inability to  agree to %
  settlement-free or paid peering.

\textbf{An Example in IPv4:}
We next explore a real-world example of partial reachability to
  several Polish ISPs.
Our algorithms found that
  on 2017-10-23, for a period of 3 hours starting at 22:02Z,
  five Polish \acp{AS}
  had 1716 /24 blocks that were unreachable from five \acp{VP},
  but they remained reachable from a sixth \ac{VP}.

Before the peninsula, all blocks that became partially unreachable
  received service through Multimedia Polska (AS21021, or \emph{MP}),
  via Cogent (AS174), with an alternate path through Tata (AS6453).
When the peninsula occurred, traffic continued through Cogent
  but was blackholed; it did not shift to Tata (see \autoref{sec:polish_peninsula_validation}).
One \ac{VP} (W) could reach MP
  through  Tata for the entire event,
  proving MP was connected.
After 3 hours, we see a burst of BGP updates (more than 23k),
  making MP reachable again from all VPs.

\begin{figure}
  \includegraphics[width=0.9\columnwidth]{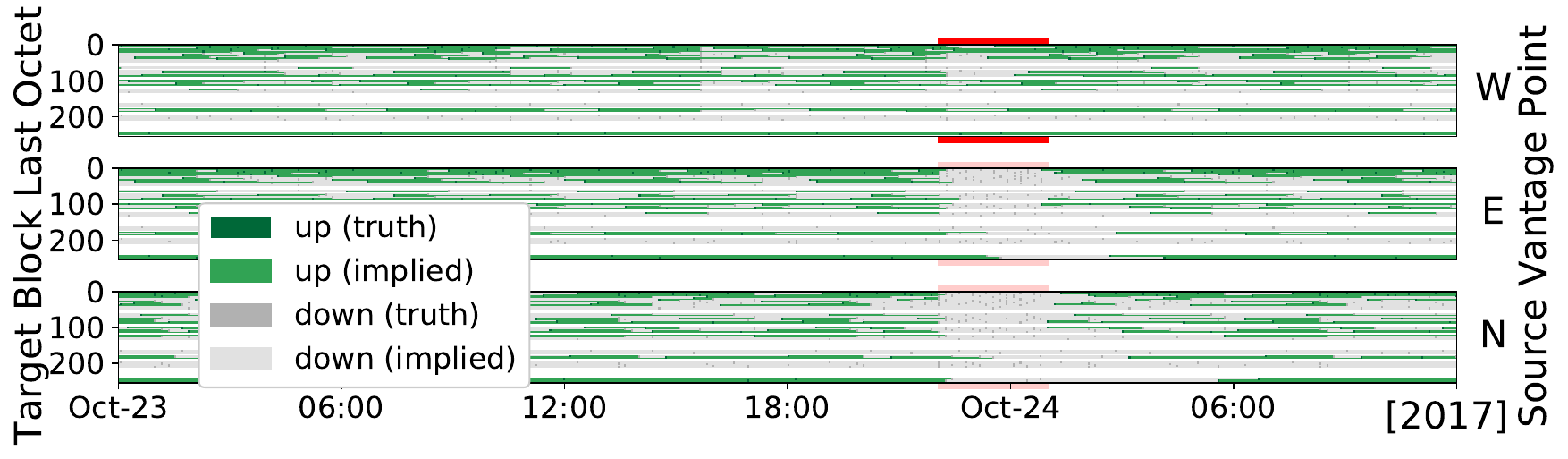}
  \caption{\small A 3-hour peninsula where
      block 80.245.176.0/24
      is reachable from \ac{VP} W (top)
    and not other \acp{VP} (E and N shown).
    (2017q4)
    }
  \label{fig:a30all_raw_50f5b000}
  \vspace*{-4ex}
\end{figure}

To show how our algorithms detect this event,
  \autoref{fig:a30all_raw_50f5b000} shows responses for one block.
In this case the top \acp{VP} always reach the block,
  but the lower two are unreachable (all address gray) for 3 hours.

We can confirm this peninsula with additional observations
  from traceroutes taken by CAIDA's Archipelago~\cite{CAIDA07b} (Ark).
During the event we see 94 unique Ark VPs attempted 345 traceroutes to the affected blocks.
Of the 94 VPs, 21 VPs (22\%) have their last responsive
  traceroute hop in the same \ac{AS} as the
  target address, and 68 probes (73\%) stopped before reaching that \ac{AS}.
The remaining 5 VPs were able to reach the destination \ac{AS} for some
  probes, while not for others.
(Sample traceroutes are in \autoref{sec:polish_peninsula_validation}.)

Although we do not have a root cause for this peninsula from network operators,
  large number of BGP Update messages suggests a routing problem.
 In \autoref{sec:peninsula_locations} we show peninsulas are mostly due to policy choices.

\section{Detecting Partial Connectivity}
	\label{sec:design}

We use observations from multiple, independent \acp{VP}
  to detect partial outages and islands (from \autoref{sec:problem})
  with our new algorithms:
\emph{Taitao} detects peninsulas,
  and \emph{Chiloe}, islands.
(Algorithm names are from Patagonian geography.)

\subsection{Suitable Data Sources}
	\label{sec:data_sources}

We use publicly available data from three systems: USC Trinocular~\cite{quan2013trinocular},
  RIPE Atlas~\cite{Ripe15c},
  and CAIDA's Archipelago~\cite{ark_data} (Ark).
We list all datasets in \autoref{tab:datasets} in \autoref{sec:data_sources_list}.

Our algorithms use data from Trinocular~\cite{quan2013trinocular}
  because it is available at no cost~\cite{LANDER14d},
  provides data since 2014,
  and covers most of the responsive IPv4
  Internet~\cite{Baltra20a}.
Briefly, Trinocular watches
  about 5M out of 5.9M responsive IPv4 /24 blocks.
In each probing round of 11 minutes,
  it sends up to 15 ICMP echo-requests (pings),
  stopping early if it proves the block is reachable.
It interprets the results using Bayesian inference,
  and merges the results from six geographically distributed \acp{VP}.
\acp{VP} are in Los Angeles (W), Colorado (C), Tokyo (J), Athens (G),
  Washington, DC (E), and Amsterdam (N).
In \autoref{sec:site_correlation} we show they are topologically independent.
Our algorithms should work with other active probing data
  as future work.

We use RIPE Atlas~\cite{Ripe15c} for islands (\autoref{sec:chiloe})
  and to see how partial connectivity affects monitoring (\autoref{sec:dnsmon}).
As of 2022, it has about 12k \acp{VP},  %
  distributed globally in over 3572 different IPv4 ASes.
Atlas VPs carry out both researcher-directed measurements
  and periodic scans of DNS servers.
We use Atlas scans of DNS root servers in our work.

We validate our results using CAIDA's Ark~\cite{ark_data},
  and use AS numbers from Routeviews~\cite{routeviews}.

We generally use recent data, but in some cases we chose older data
  to avoid known problems in measurement systems.
Many of our findings are demonstrated over multiple years,
  as we show in \autoref{sec:2020}.
We use Trinocular measurements for 2017q4 because
  this time period had six active VPs,
  allowing us to make strong statements about how multiple perspectives help.
We use 2020q3 data in
  \autoref{sec:peninsula_locations}
  because Ark observed a very large number of loops in 2017q4.
Problems with different VPs reduced coverage for 2019 and 2020,
  but we verify and find quantitatively similar results in 2020 in \autoref{sec:2020}).

\subsection{Taitao: a Peninsula Detector}
\label{sec:disagreements}

Peninsulas occur when portions of the Internet
  are reachable from some locations and not others.
They can be seen by two \acp{VP} disagreeing on reachability.
With multiple VPs, non-unanimous observations
  suggest a peninsula.

Detecting peninsulas presents three challenges.
First, we do not have \acp{VP} everywhere.
If all \acp{VP} are on the same ``side'' of a peninsula,
  their reachability agrees even though other potential \acp{VP} may disagree.
Second, \ac{VP} observations are not synchronized.
For Trinocular, they are spread over an 11-minute interval,
  so different \acp{VP} test reachability
  at slightly different times.
When observations are made just before and after a network change,
  both are true but the disagreement is from unsynchronized measurement
  and not a peninsula.
Third, connectivity problems near the observer (or when an observer is an island)
  should not reflect on the intended destination.

We identify peninsulas by detecting disagreements in
  block state by comparing
  valid \ac{VP} observations that occur at about the same time.
Since probing rounds occur every 11 minutes,
  we compare measurements within an 11-minute window.
This approach will see peninsulas that last at least 11 minutes,
  but may miss briefer ones,
  or peninsulas where \acp{VP} are not on ``both sides''.

Formally, $O_{i,b}$ is the set of observers with valid observations
about block $b$ at round $i$.
We look for disagreements in $O_{i,b}$,
  defining $O_{i,b}^\Vs{up} \subset O_{i,b}$ as the
set of observers that measure block $b$ as up at round $i$.
We detect a peninsula when:
\begin{align}
  0 < |O_{i,b}^\Vs{up}| < |O_{i,b}|
\end{align}

When only one \ac{VP} reaches a block,
  that block can be either a peninsula or an island.
We require more information to distinguish them,
  as we describe
 in \autoref{sec:chiloe}.

\subsection{Detecting Country-Level Peninsulas}
\label{sec:detecting_country_peninsulas}

Taitao detects peninsulas based on differences in observations.
Long-lived peninsulas are likely intentional, from policy choices.
One policy is filtering based on national boundaries,
  possibly to implement legal requirements about data sovereignty
  or economic boycotts.

We identify country-specific peninsulas as a special case of Taitao
  where a given destination block is reachable (or unreachable) from only one country,
  persistently for an extended period of time.
(In practice,
  the ability to detect country-level peninsulas is somewhat limited because
  the only country with multiple VPs in our data is the United States.
However, we augment non-U.S.~observers with data from other non-U.S.~sites
  such as Ark or RIPE Atlas.)

A country level peninsula occurs when
  \emph{all} available \acp{VP} from the same country as the target block
  successfully reach the target block and all available \acp{VP} from different countries
  fail.
Formally, we say there is a country peninsula when the set of observers claiming
block $b$ is up at time $i$ is equal to $O_{i,b}^c \subset O_{i,b}$
the set of all available observers with
valid observations at country $c$.
\begin{align}
  O_{i,b}^\Vs{up} = O_{i,b}^c
\end{align}

\subsection{Chiloe: an Island Detector}
	\label{sec:chiloe}

According to our definition in \autoref{sec:island}, islands occur
when the Internet is partitioned, and the smaller component
  (that with less than half the active addresses)
  is the island.
Typical islands are much, much smaller.

We can find islands by looking for networks that
  are only reachable from less than half of the Internet.
However, to classify such networks as an island
  and not merely a peninsula, we need to show that it is partitioned.
Without global knowledge, it is difficult to prove disconnection.
In addition, if islands are partitioned from \acp{VP},
  we cannot tell an island from an outage.
An island is disconnected
  but still active inside,
  but for an outage, the computers are disconnected from the Internet's core
  \emph{and} from each other.

For these reasons, %
  we must look for islands that include \acp{VP} in their partition.
Because we know the VP is active and scanning
  we can determine how much of the Internet is in its partition,
  ruling out an outage.
We also can confirm the Internet is not reachable,
  to rule out a peninsula.

Formally, we say that $B$ is the set of all blocks on the Internet
  responding in the last week.
$B^\Vs{up}_{i,o} \subseteq B$ are blocks reachable from observer
$o$ at round $i$, while
$B^\Vs{dn}_{i,o} \subseteq B$ is its complement.
We detect that observer $o$ is in an island when
  it thinks half or more of the
  observable Internet
  is down:
\begin{align}
  0 \leq |B^\Vs{up}_{i,o}| \leq |B^\Vs{dn}_{i,o}|
\end{align}
This method is independent from measurement systems,
  but is limited to detecting islands that contain \acp{VP}.
We evaluate islands in two systems with thousands of VPs in
  \autoref{sec:how_common_are_islands}.
Finally, because observations are not instantaneous,
  we must avoid confusing short-lived islands with long-lived peninsulas.
For islands lasting longer than an observation period, we also require
$|B^\Vs{up}_{i,o}| \rightarrow 0$.
When $|B^\Vs{up}_{i,o}| = 0$, then we have an address island.

\subsection{Applications}
        \label{sec:applications}
	\label{sec:who_has_the_internet}

\textbf{Political: Who Has the Internet?}  We explore this question in
\autoref{sec:other_applications} and
\autoref{sec:internet_partition}.

\textbf{Architectural:} Our work helps understand risk
  by showing reachability is not binary,
  but often partial.
We explore this issue in \autoref{sec:evaluation};
  one key result is that users see peninsulas as often as outages (\autoref{sec:peninsula_frequency}).
It helps clarify
  prior studies of Internet outages~\cite{Schulman11a,quan2013trinocular,Shah17a,richter2018advancing,guillot2019internet} (more detail is in \autoref{sec:local_outage_eval}).

\textbf{Operational: Cleaning Data.}
Problems near network observers can skew observations
  and must be detected and removed,
  as we explore in \autoref{sec:dnsmon}
  and~\cite{Saluja22a} and detection of Covid-work-from-home~\cite{Song21a}.

\section{Validating our approach}
	\label{sec:validation}

We validate our algorithms,
  comparing Taitao peninsulas
  and Chiloe islands
  to independent data (\autoref{sec:taitao_validation} and
  \autoref{sec:chiloe_validation})%
, and examining country-level peninsulas~(\autoref{sec:country_validation})%
.

\begin{table*}
  \begin{minipage}[b]{.31\linewidth}
  \footnotesize
  \resizebox{\textwidth}{!}{
  \begin{tabular}{c c c | c c c}
    & & & \multicolumn{3}{c}{\normalsize \textbf{Ark}} \\
    & & Sites Up & Conflicting & All Down & All Up \\
	\cline{3-6}
    \multirow{7}{2pt}{\rotatebox[origin=l]{90}{\parbox{1.6cm}{\normalsize \textbf{Trinocular}}}}
    & \multirow{5}{2pt}{\rotatebox[origin=l]{90}{\parbox{35pt}{Conflicting}}}
      & 1 & \cellcolor[HTML]{99ee77}20  & \cellcolor[HTML]{99ee77}6  & \cellcolor[HTML]{FFF9C4}\emph{15}  \\
    & & 2 & \cellcolor[HTML]{99ee77}13  & \cellcolor[HTML]{99ee77}5  & \cellcolor[HTML]{FFF9C4}\emph{11}  \\
    & & 3 & \cellcolor[HTML]{99ee77}13  & \cellcolor[HTML]{99ee77}1  & \cellcolor[HTML]{FFF9C4}\emph{5}   \\
    & & 4 & \cellcolor[HTML]{99ee77}26  & \cellcolor[HTML]{99ee77}4  & \cellcolor[HTML]{FFF9C4}\emph{19}  \\
    & & 5 & \cellcolor[HTML]{99ee77}83  & \cellcolor[HTML]{99ee77}13 & \cellcolor[HTML]{FFF9C4}\emph{201} \\
	\cline{3-6}
    & \multirow{2}{2pt}{\rotatebox[origin=l]{90}{\parbox{21pt}{Agree}}}
    & 0 & \cellcolor[HTML]{F0ABAB}\textbf{6} & \cellcolor[HTML]{CCFF99}97 & \cellcolor[HTML]{F0ABAB}\textbf{6}     \\
    & & 6 & \cellcolor[HTML]{CCFF99}491,120 & \cellcolor[HTML]{CCFF99}90
    & \cellcolor[HTML]{CCFF99}1,485,394 \\
\end{tabular}}
  \caption{Trinocular and Ark agreement table. Dataset A30, 2017q4.}
  \label{tab:taitao_validation_table}
\end{minipage}
\hspace{3mm}
\begin{minipage}[b]{.31\linewidth}
  \centering
  \footnotesize
  \resizebox{\textwidth}{!}{
  \begin{tabular}{c P{40pt} P{40pt} c c}
    & & \multicolumn{3}{c}{\normalsize \textbf{Ark}} \\
    & & Peninsula & \multicolumn{2}{c}{Non Peninsula} \\
    \multirow{2}{2pt}{\rotatebox[origin=l]{90}{\parbox{30pt}{\normalsize \textbf{Taitao}}}}
    & Peninsula & \cellcolor[HTML]{99ee77} 184 & \cellcolor[HTML]{FFF9C4}
    \emph{251 (strict)} & \parbox[30pt][20pt][c]{30pt}{\cellcolor[HTML]{FDD835} \emph{40 (loose)}} \\
    & \parbox[40pt][20pt][c]{40pt}{\centering Non Peninsula} & \cellcolor[HTML]{F0ABAB} \textbf{12} &
    \multicolumn{2}{c}{\cellcolor[HTML]{CCFF99} 1,976,701} \\
  \end{tabular}
}
  \vspace{10pt}
  \caption{Taitao confusion matrix.  Dataset A30, 2017q4.}
  \label{tab:taitao_confusion_matrix}
\end{minipage}
\hspace{3mm}
  \begin{minipage}[b]{.31\linewidth}
  	\footnotesize
  	\tabcolsep=0.1cm
  	\renewcommand{\arraystretch}{1.1}
\resizebox{\textwidth}{!}{
    \begin{tabular}{c c | c c c | c}
      & & \multicolumn{3}{c}{\normalsize \textbf{Ark}} & \\
      & U.S. VPs & Domestic Only & $\leq5$ Foreign & $>5$ Foreign & Total\\
  	\cline{2-6}
      \multirow{7}{2pt}{\rotatebox[origin=l]{90}{\parbox{50pt}{\centering \normalsize \textbf{Trinocular}}}}
      & WCE & \cellcolor[HTML]{99ee77}211  & \cellcolor[HTML]{99ee77}171  & \cellcolor[HTML]{FFF9C4}\emph{47} & 429  \\
      & WCe & \cellcolor[HTML]{F0ABAB}\textbf{0}  & \cellcolor[HTML]{CCFF99}5  & \cellcolor[HTML]{CCFF99}1 & 6  \\
      & WcE & \cellcolor[HTML]{F0ABAB}\textbf{0}  & \cellcolor[HTML]{CCFF99}1  & \cellcolor[HTML]{CCFF99}0 & 1   \\
      & wCE & \cellcolor[HTML]{F0ABAB}\textbf{0}  & \cellcolor[HTML]{CCFF99}0  & \cellcolor[HTML]{CCFF99}0 & 0  \\
      & Wce & \cellcolor[HTML]{F0ABAB}\textbf{3}  & \cellcolor[HTML]{CCFF99}40 & \cellcolor[HTML]{CCFF99}11 & 54\\
      & wcE & \cellcolor[HTML]{F0ABAB}\textbf{0} & \cellcolor[HTML]{CCFF99}4 & \cellcolor[HTML]{CCFF99}5    & 9     \\
      & wCe & \cellcolor[HTML]{F0ABAB}\textbf{0} & \cellcolor[HTML]{CCFF99}1 & \cellcolor[HTML]{CCFF99}1    & 2 \\
      \midrule
      & Marginal distr.  & 214 & 222 & 65 & 501 \\
  \end{tabular}}
    \caption{Trinocular U.S.-only blocks. Dataset A30, 2017q4.}
    \label{tab:taitao_countries_validation_table}
  \end{minipage}
\end{table*}

\subsection{Can Taitao Detect Peninsulas?}
\label{sec:taitao_validation}

We compare Taitao detections
  from 6 \acp{VP}
 to independent observations taken from more than 100 VPs in CAIDA's Ark~\cite{ark_data}.
This comparison is challenging,
  because both Taitao and Ark are imperfect operational systems
  that differ in probing frequency, targets, and method.
Neither defines perfect ground truth,
  but agreement suggests likely truth.

Although Ark probes targets much less frequently than Trinocular,
  Ark makes observations from 171 global locations,
  providing a diverse perspective.
Ark traceroutes also allow us to assess \emph{where} peninsulas begin.
We expect to see a strong correlation between Taitao peninsulas and Ark observations.
(We considered RIPE Atlas as another external dataset,
  but its coverage is sparse, while Ark covers all /24s.)

\textbf{Identifying comparable blocks:}
We study 21 days of Ark observations from 2017-10-10 to -31.
Ark covers all networks with two strategies.
With team probing,
  a 40 VP ``team'' traceroutes to
  all routed /24 about once per day.
For prefix probing,
  about 35 VPs each traceroute to .1 addresses of all routed /24s every day.
We use both types of data: the three Ark teams
  and all available prefix probing VPs.
We group results by /24 block of the traceroute's target address.

Ark differs from Taitao's Trinocular input in three ways:
  the target is a random address or the .1 address in each block;
  it uses traceroute, not ping;
  and it probes blocks daily, not every 11 minutes.
Sometimes these differences cause Ark traceroutes to fail
  when a simple ping succeeds.
First, Trinocular's targets respond more often because
  it uses a curated hitlist~\cite{Fan10a}
  while Ark does not.
Second, Ark's traceroutes can terminate due to path
  \emph{loops}
  or \emph{gaps} in the path,
  (in addition to succeeding or reporting target unreachable).
We do not consider results with gaps, %
  so problems on the path do not
  bias results for endpoints reachable by direct pings.

To correct for differences in target addresses,
  we must avoid
  misinterpreting a block as unreachable
  when the block is online but Ark's target address is not,
  we discard traces sent to never-active addresses
  (those not observed in 3 years of complete IPv4 scans), %
  and blocks
  for which Ark did not get a single successful response.%
(Even with this filtering,
  dynamic addressing means Ark still sometimes sees unreachables.)

To correct for Ark's less frequent probing,
  we compare \emph{long-lived} Trinocular down-events (5 hours or more).
Ark measurements are infrequent (once every 24 hours) compared to Trinocular's 11-minute reports,
  so short Trinocular events are often unobserved by Ark.
To confirm agreements or conflicting reports from Ark,
  we require at least 3 Ark observations within the peninsula's span of time.

We filter out blocks with frequent transient changes
  or signs of network-level filtering.
We define the ``reliable'' blocks suitable for comparison
  as those responsive for at least 85\% of the quarter
  from each of the 6 Trinocular VPs.
(This threshold avoids diurnal blocks or blocks with long outages;
  values of 90\% or less have similar results.)
We also discard flaky blocks whose responses are frequently inconsistent across \acp{VP}.
(We consider more than 10 combinations of \ac{VP} as frequently inconsistent.)
For the 21 days, we find 4M unique Trinocular /24 blocks,
  and 11M Ark /24 blocks,
  making 2M blocks in both available for study.

\textbf{Results:}
\autoref{tab:taitao_validation_table} provides details
  and \autoref{tab:taitao_confusion_matrix} summarizes our interpretation.
Here dark green indicates true positives (TP):
  when (a) either both Taitao and Ark show mixed results, both indicating a peninsula,
  or when (b) Taitao indicates a peninsula (1 to 5 sites up but at least one down),
Ark shows all-down during the event and up before and after.
We treat Ark in case (b) as positive
  because the infrequency of Ark probing
  (one probe per team every 24 hours) %
  means we cannot guarantee
  VPs in the peninsula will probe responsive targets in time.
Since peninsulas are rare, so too are true positives,
  but we see 184 TPs.

We show \emph{true negatives} as light green and neither bold nor italic.
In almost all of these cases (1.4M)
  both Taitao and Ark both reach the block, agreeing.
Because of dynamic addressing~\cite{Padmanabhan16a},
  many Ark traceroutes end in a failure at the last hop
  (even after we discard never-reachable).
We therefore count this second most-common result (491k cases) as a true negative.
For the same reason, we include the small number (97) of cases where
  Ark reports conflicting results and Taitao is all-up,
  assuming Ark terminates at an empty address.
We include in this category the 90 events where Ark is all-down and Trinocular
is all-up.
We attribute Ark's failure to reach its targets to infrequent probing.

We mark \emph{false negatives} as red and bold.
For these few cases (only 12),
  all Trinocular \acp{VP} are down, but
  Ark reports all or some responding.
We believe these cases indicate blocks that have chosen to drop Trinocular traffic.

Finally, yellow italics shows cases where a Taitao peninsula
  is a \emph{false positive}, since all Ark probes reached the target block.
This case occurs when either traffic from some Trinocular \acp{VP} is
  filtered, or all Ark VPs are ``inside'' the peninsula.
Light yellow (strict) shows all the 251 cases that Taitao detects.
For most of these cases (201),
  five Trinocular \acp{VP} responding and one does not,
  suggesting network problems are near one of the Trinocular VPs
  (since five of six independent VPs have working paths).
Discarding these cases we get 40 (orange); still conservative but a \emph{looser} estimate.

The strict scenario sees precision 0.42, recall 0.94, and $F_1$ score
  0.58,
  and in the loose scenario, precision improves to 0.82 and
  $F_1$ score to 0.88.
We consider these results good, but with some room for improvement.

\subsection{Detecting Country-Level Peninsulas}
	\label{sec:country_validation}

Next, we verify detection of country-level peninsulas
(\autoref{sec:detecting_country_peninsulas}).
We expect that legal requirements sometimes result in long-term
  network unreachability.
For example, blocking access from Europe
  is a crude way to comply with the EU's GDPR~\cite{eu_gdpr}.

Identifying country-level peninsulas requires
  multiple VPs in the same country.
Unfortunately the source data we use only has multiple \acp{VP} for the United States.
We therefore look for U.S.-specific peninsulas
  where only these \acp{VP} can reach the target and the non-U.S.-\acp{VP} cannot,
  or vice versa.

We first consider the 501 cases where Taitao reports that only U.S.~\acp{VP}
  can see the target, and compare to how Ark \acp{VP} respond.
For Ark, we follow
\autoref{sec:taitao_validation},
  except retaining blocks with less than 85\% uptime.
We only consider Ark VPs that are able
  to reach the destination (that halt with ``success'').
We note blocks that can only be reached by Ark VPs within the same
country as domestic, and blocks that can be reached from VPs located in other
countries as foreign.

In \autoref{tab:taitao_countries_validation_table}
  we show the number of blocks that uniquely
  responded to all U.S.~\ac{VP} combinations during the quarter.
We contrast these results against Ark reachability.

True positives are when  Taitao shows a peninsula
  responsive only to U.S.~\acp{VP}
  and nearly all Ark \acp{VP} confirm this result.
We see 211 targets are U.S.-only, and another 171 are available to only a few
  non-U.S.~countries.
The specific combinations vary: sometimes allowing access from the U.K.,
  or Mexico and Canada.
Together these make 382 true positives, most of the 501 cases.
Comparing all positive cases, we see a
  very high precision of 0.99 (382 green of 385 green and red reports)---our
  predictions are nearly all confirmed by Ark.

In yellow italics we show 47 cases of false positives
  where more than five non-U.S. countries are allowed access.
In many cases these include many European countries.
Our recall is therefore 0.89 (382 green of 429 green and yellow true country peninsulas).

In light green we show true negatives.
Here we include blocks that filter one or more U.S. \acp{VP},
and are reachable from Ark VPs in multiple
countries, amounting to a total of 69 blocks.
There are other categories involving non-U.S. sites,
  along with other millions of true
  negatives, however, we only concentrate in these few.

In red and bold we show three false negatives.
These three blocks seem to have strict filtering policies,
  since they were reachable only from one U.S.~site (W)
  and not the others (C and E) in the 21 days period.

\subsection{Can Chiloe Detect Islands?}
\label{sec:chiloe_validation}

Chiloe (\autoref{sec:chiloe}) detects islands when a \ac{VP} within the island
  can reach less than half the rest of the world.
When less than 50\% of the network replies,
  it means that the \ac{VP} is either in an island (for brief events, or when replies drop near zero)
  or a peninsula (long-lived partial replies).

To validate Chiloe's correctness,
  we compare when a single \ac{VP} believes to be in an island,
  against what the rest of the world believes about that \ac{VP}.

We define ground truth at a block level granularity---if \ac{VP} $x$ can reach its own block when $x$ believes to be
in an island,
while other external \acp{VP} can't reach $x$'s block,
then $x$'s island is confirmed.
On the other hand, if an external \ac{VP} can reach $x$'s block, then
  $x$ is not in island, but in a peninsula.
In \autoref{sec:site_correlation} we show that Trinocular \acp{VP} are independent,
  and therefore no two \acp{VP} live within the same island.
We believe this definition is the best possible ground truth,
  but of course a perfect identification of island or peninsula requires instant,
  global knowledge and so cannot be measured in practice.

We take 3 years worth of data from all six
  Trinocular \acp{VP}.
Because Trinocular spreads measurements over 11 minutes,
  we group results into 11-minute bins.

  \begin{table}
\centering
	\resizebox{0.49\columnwidth}{!}{
\subfloat[Chiloe confusion matrix \label{tab:chiloe_validation}]{%
    \centering
    \begin{tabular}{c c P{30pt} P{35pt} @{}m{0pt}@{}}
      & & \multicolumn{2}{c}{\normalsize \textbf{Chiloe}} \\
      & & \parbox{40pt}{\centering Island} & \parbox{40pt}{\centering Peninsula} \\
      \multirow{3}{10pt}{\rotatebox[origin=l]{90}{\parbox{50pt}{\normalsize \textbf{Trinocular}}}} &
      \parbox[10pt][20pt][c]{40pt}{Blk Island} &
      \cellcolor[HTML]{99ee77} 2 & \cellcolor[HTML]{F0ABAB} \textbf{0} \\
      & \parbox[10pt][20pt][c]{50pt}{Addr Island} &
      \cellcolor[HTML]{99ee77} 19 & \cellcolor[HTML]{F0ABAB} \textbf{8} \\
      & \parbox[10pt][20pt][c]{40pt}{Peninsula}
      & \cellcolor[HTML]{FFF9C4} \emph{2} & \cellcolor[HTML]{CCFF99} 566
    \end{tabular}

	  }}%
\qquad%
	\resizebox{0.42\columnwidth}{!}{
\subfloat[Detected islands\label{tab:island_summary}]{%
    \centering
  \begin{tabular}{c c c}
   Sites	& Events	& Per Year \\
   \midrule
   W	    & 5		    & 1.67 \\
   C	    & 11  		& 3.67 \\
   J	    & 1		    & 0.33 \\
   G	    & 1		    & 0.33 \\
   E	    & 3		    & 1.00 \\
   N	    & 2		    & 0.67 \\
    \hline
    All (norm.)     & 23        & 7.67 (1.28)  \\
  \end{tabular}

	  }}
    \caption{(a) Chiloe confusion matrix, events between 2017-01-04 and 2020-03-31, datasets A28 through A39.
(b) Islands detected from 2017q2 to 2020q1.
	  }
\vspace*{-2ex}
\end{table}

In \Cref{tab:chiloe_validation} we show that Chiloe detects 23 islands
across three years.
In 2 of these events, the block is unreachable from other \acp{VP},
confirming the island with our ground-truth methodology.
Manual inspection confirms that the remaining
19 events are islands too, but at the address level---the \ac{VP}
  was unable to reach anything but did not lose power,
  and other addresses in its block were reachable from \acp{VP} at other locations.
These observations suggest a VP-specific problem making it an island.
Finally, for 2 events, the prober's block was reachable during the event by
every site including the prober itself which suggests partial connectivity
(a peninsula), and therefore a false positive.

In the 566 non-island events (true negatives),
  a single \ac{VP} cannot reach more than 5\% but less than 50\% of
  the Internet core.
In each of these cases, one or more
other \acp{VP} were able to reach the affected \ac{VP}'s block,
  showing they were not an island (although perhaps a peninsula).
We omit the very frequent events when less than 5\% of the network is unavailable from the \ac{VP}
  from the table, although they too are true negatives.

Bold red shows 8 false negatives. These are events that last about 2 Trinocular
rounds or less (22 min), often not enough time for Trinocular to change its
belief on block state.

\subsection{Are the Sites Independent?}
	\label{sec:site_correlation}

Our evaluation assumes \acp{VP} do not share common network paths.
Two \acp{VP} in the same location would share the same local outages,
  but those in different physical locations
  will often use different network paths,
  particularly with a ``flatter'' Internet graph~\cite{Labovitz10c}.
We next quantify this similarity to validate our assumption.

We next measure similarity of observations
  between pairs of VPs.
We examine only cases where one of the pair disagrees with some other VP,
  since when all agree, we have no new information.
If the pair agrees with each other, but not with the majority,
  the pair shows similarity.
If they disagree with each other, they are dissimilar.
We quantify similarity $S_P$ for a pair of sites $P$ as
	${S_P = (P_1 + P_0)/(P_1 + P_0 + D_*)}$,
where $P_s$ indicates the pair agrees on the network having state $s$ of
  up (1) or down (0) and disagrees with the others,
  and for $D_*$, the pair disagrees with each other.
$S_P$ ranges from 1, where the pair always agrees,
  to 0, where they always disagree.

\autoref{tab:overall_correlation}(a) shows similarity values for each pair
of the 6 Trinocular VPs.
(We show only half of the symmetric matrix.)
No two sites have a similarity more than 0.14,
  and most pairs are under 0.08.
This result shows that no two sites are particularly correlated.

\section{Internet Islands and Peninsulas}
	\label{sec:evaluation}

We now examine islands and peninsulas in the Internet core.

\subsection{How Common are Peninsulas?}
	\label{sec:peninsula_frequency}

We estimate how common peninsulas occur in the Internet core
  in three ways.
First, we directly measure the visibility of peninsulas in the Internet
  by summing the duration of peninsulas as seen from six VPs.
Second, we confirm the accuracy of this estimate
  by evaluating its convergence as we vary the number of VPs---more VPs
  show more peninsula-time, but if the result converges we predict
  we are approaching the limit.
Third, we compare peninsula-time to outage-time,
  showing that, in the limit, observers see both for about the same
  duration.
Outages correspond to service downtime~\cite{down_time_cost},
  and are a recognized problem in academia and industry.
Our results show that \emph{peninsulas are as common as outages},
  suggesting peninsulas are an important new problem deserving attention.

\textbf{Peninsula-time:}
We estimate the duration an observer can see a peninsula
  by considering three types of events: \emph{all up}, \emph{all down}, and
  \emph{disagreement} between six VPs.
Disagreement, the last case, suggests a peninsula,
  while agreement (all up or down),
  suggests no problem or an outage.
We compute peninsula-time by summing the time each target /24
  has disagreeing observations from Trinocular VPs.

\begin{figure*}
\adjustbox{valign=b}{\begin{minipage}[b]{.25\linewidth}
    \includegraphics[width=1\linewidth]{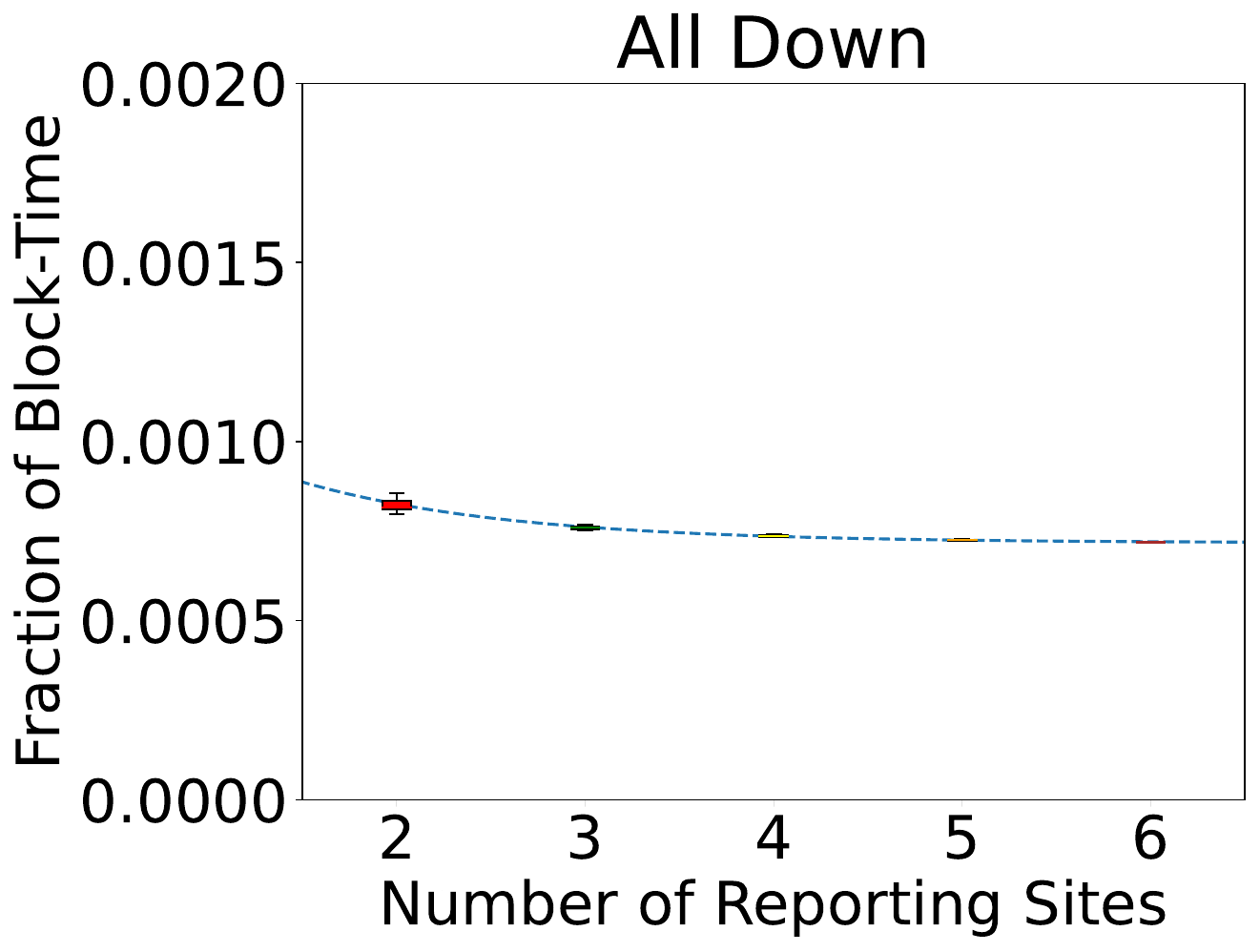}
\end{minipage}}\quad
\adjustbox{valign=b}{\begin{minipage}[b]{.25\linewidth}
    \includegraphics[width=1\linewidth]{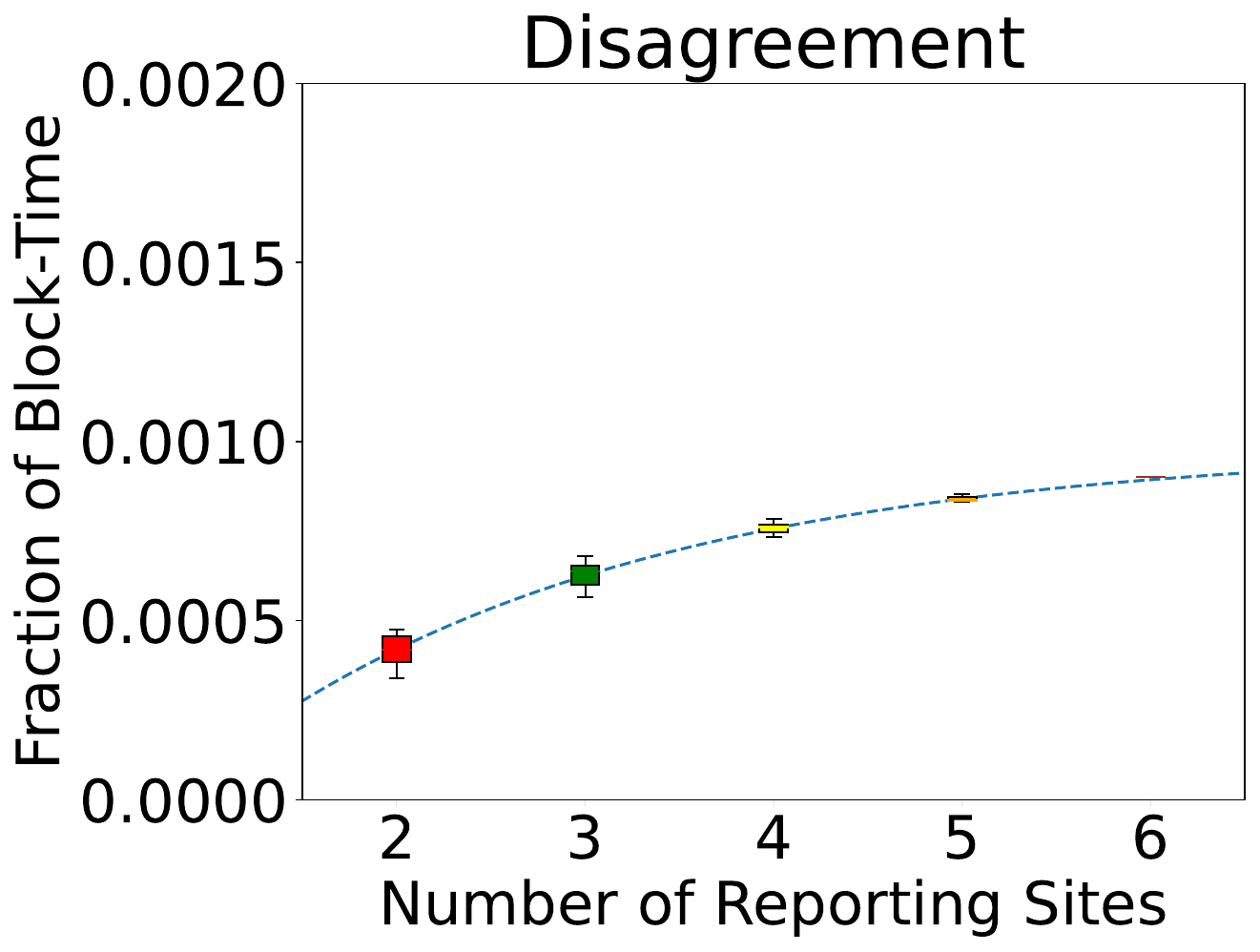}
\end{minipage}}\quad
\adjustbox{valign=b}{\begin{minipage}[b]{.32\linewidth}
    \includegraphics[width=1\linewidth]{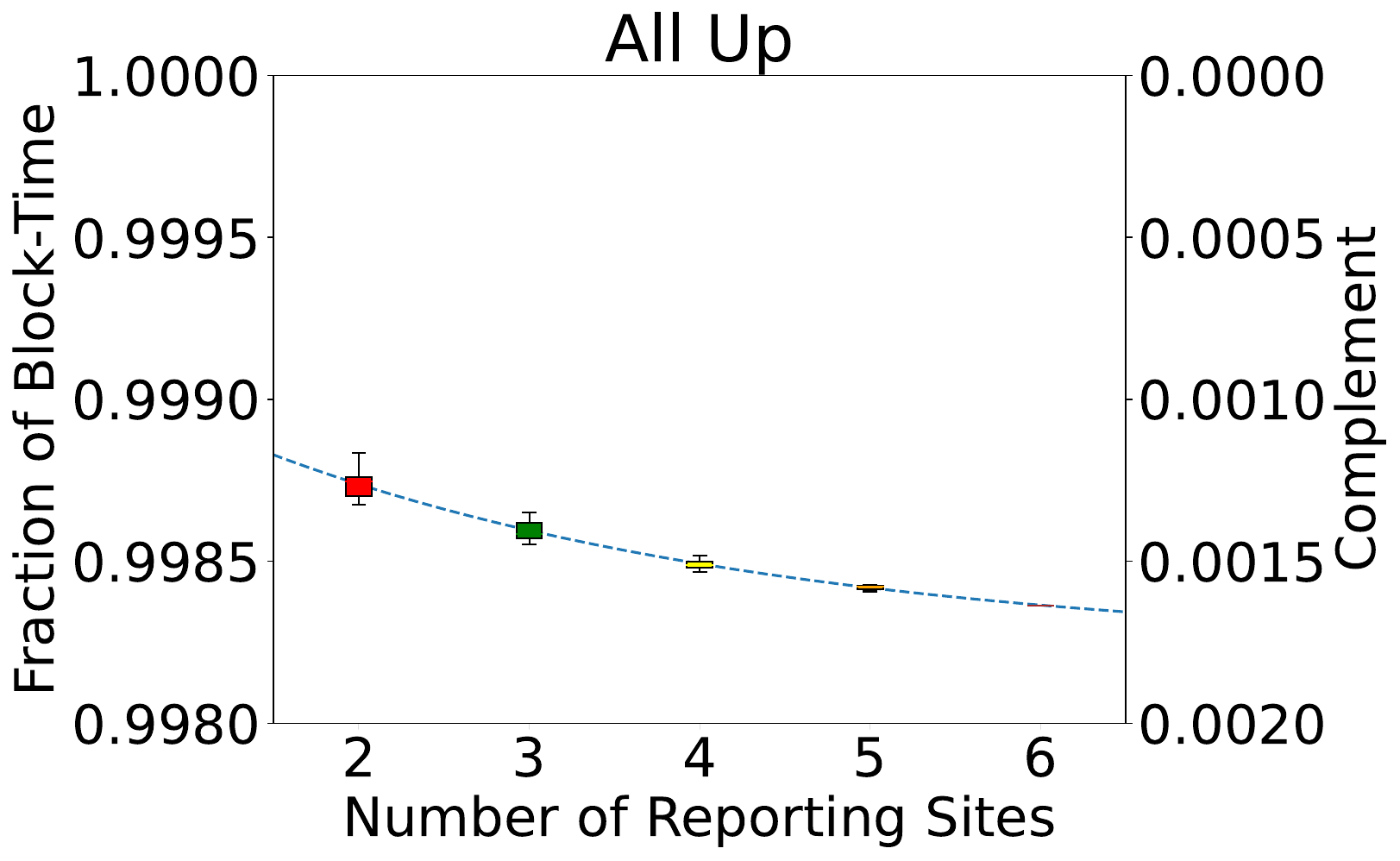}
\end{minipage}}
\caption{Distribution of block-time fraction for 3.7M blocks over sites reporting all down
(left), disagreement (center), and all up (right), for events longer than one
hour. Dataset A30, 2017-10-06 to 2017-11-16.}
\label{fig:a30all_peninsulas_duration_oct_nov}
\end{figure*}

We have computed peninsula-time
  by evaluating Taitao over Trinocular data for 2017q4~\cite{LANDER14d}.
\autoref{fig:a30all_peninsulas_duration_oct_nov} shows the distribution of peninsulas
measured as a fraction of block-time for an increasing number of sites.
We consider all possible combinations of the six sites.

First we examine the data with all 6 VPs---the rightmost point on each graph.
We see that %
  peninsulas (the middle, disagreement graph)
  are visible about 0.00075 of the time.
This data suggests \emph{peninsulas are rare, occurring less than 0.1\% of the time,
  but do regularly occur}.

\textbf{Convergence:}
With more \acp{VP} we get a better view of the Internet's overall state.
As more reporting sites are added, more peninsulas are discovered.
That is previously inferred outages (all unreachable) should have been peninsulas,
  with partial reachability.
All-down (left) decreases from an average of 0.00082 with 2 \acp{VP} to 0.00074 for 6
\acp{VP}. All-up (right) goes down a relative 47\% from 0.9988 to 0.9984, while
disagreements (center) increase from 0.0029 to 0.00045.
\emph{Outages (left) converge after 3 sites},
  as shown by the fitted curve and decreasing variance.
Peninsulas and all-up converge more slowly.
We conclude that \emph{a few sites (3 or 4) converge on a good estimate of true islands
  and peninsulas, provided they are independently located}.

We can support this claim by comparing
  all non-overlapping combinations of 3 sites.
If any combination is equivalent with any other,
  then a fourth site would not add new information.
There are 10 possible pairs of 3 sites from 6 observers,
  and we examine those combinations for each of 21 quarters, from 2017q2 to 2020q1.
When we compare the one-sample Student $t$-test
  to evaluate if the difference of each pair of combinations of those 21 quarters
  is greater than zero.
None of the combinations are rejected at confidence level 99.75\%,
  suggesting that any combination of three sites is statistically equivalent
  and confirm our claim that a few sites are sufficient for estimation.

\textbf{Relative impact:}
Finally, comparing outages (the left graph) with peninsulas (the middle graph),
  we see both occur about the same fraction of time (around 0.00075).
This comparison shows that \emph{peninsulas are about as common as outages},
  suggesting they deserve more attention.

\textbf{Generalizing:}
We confirm these results with other quarters
  in \autoref{sec:2020}.
While we reach a slightly different limit (in that case,
  peninsulas and outages appear about in 0.002 of data),
  we still see good convergence after 4 VPs.

\subsection{How Long Do Peninsulas Last?}
	\label{sec:peninsula_duration}

Peninsulas have multiple root causes:
  some are short-lived routing misconfigurations
  while others may be long-term disagreements in routing policy.
In this section we determine the distribution of peninsulas in terms of their duration
  to determine the prevalence of persistent peninsulas.
We will show that there are millions of brief peninsulas,
  likely due to transient routing problems,
  but that 90\% of peninsula-time is in long-lived events (5\,h or more, following~\autoref{sec:taitao_validation}).

To see peninsula duration
  we use Taitao to detect peninsulas that occurred during 2017q4.
For \emph{all} peninsulas, we see 23.6M peninsulas affecting 3.8M unique blocks.
If instead we look at \emph{long-lived} peninsulas (at least 5\,h),
  we see 4.5M peninsulas in 338k unique blocks.

\autoref{fig:a30_partial_outages_duration_cdf} examines the duration of these peninsulas
  in three ways:
  the cumulative distribution of the number of peninsulas for all events
  (left, solid, purple line),
  the cumulative distribution of the number of peninsulas for VP down events
  longer than 5 hours (middle, solid green line),
  and the cumulative size of peninsulas
  for VP down events longer than 5 hours (right, green dashes).

We see that there are many very brief peninsulas (purple line).
About 65\% last from 20 to 60 minutes (about 2 to 6 measurement rounds).
Such events are not just one-off loss, since they last
  at least two observation periods.
These results suggest that while the Internet is robust,
there are many small connectivity glitches (7.8M events).
Events that are two rounds (20 minutes) or shorter
  may be due to BGP-induced transient blackholes or  measurement packet loss.

The number of day-long or multi-day peninsulas is small,
  only 1.7M events (2\%, the purple line).
However, about 57\% of all peninsula-time is in such longer-lived events
  (the right, dashed line),
  and 20\% of time is in events lasting 10 days or more,
  even when longer than 5 hours events are less numerous (compare the middle, green line to the left, purple line).
Events lasting a day last long enough that they can be debugged by human network operators,
  and events lasting longer than a week suggest potential
  policy disputes and \emph{intentional} unreachability.
Together, these long-lived events suggest that
  there is benefit to identifying non-transient peninsulas
  and addressing the underlying routing problem.

\subsection{Additional Peninsula Results}

We summarize findings omitted due to space (more in \autoref{sec:additional_peninsula_results}).

We evaluate \emph{peninsula size}
  measured as a fraction of the affected routable prefix.
We find that
  a third of peninsulas
  are much smaller than their public, routable prefix.
This evaluation suggests that peninsulas often
  happen \emph{inside} an ISP and
  are not due to interdomain routing.
Further,
  20\% of all peninsula-time
  is due to peninsulas covering their
  full routable prefixes, suggesting
  that \emph{longer-lived peninsulas are likely due to routing or policy choices} (\autoref{sec:peninsula_size}).

We also use traceroutes to estimate peninsula size.
We detect where the Internet breaks into peninsulas,
  by looking at traceroutes that failed to reach their target address, and find
  more traceroutes halt at or inside the target AS,
  but they more often terminate before reaching the target prefix.
This result suggests policy is implemented at or inside ASes, but not at routable prefixes.
By contrast, outages
  more often terminate before reaching the target AS.
Because peninsulas are more often at or in an AS,
  while outages occur in many places,
  it suggests that peninsulas are policy choices (\autoref{sec:peninsula_locations}).

Country-specific filtering is a routing policy made by networks to
restrict traffic they receive.
We next look into  what type of organizations actively block overseas
traffic.
For example, good candidates to restrain who can reach them for security purposes
  are government related organizations.

We test for country-specific filtering (\autoref{sec:detecting_country_peninsulas}) over a quarter and find 429
unique U.S.-only blocks in 95 distinct ASes confirming that,
  while not common, country specific blocks do occur (\autoref{sec:country_peninsulas}).

\subsection{How Common Are Islands?}
	\label{sec:how_common_are_islands}

Multiple groups have shown that there are many network outages in the Internet~\cite{Schulman11a,quan2013trinocular,Shah17a,richter2018advancing,guillot2019internet}.
We have described (\autoref{sec:problem}) two kinds of outages:
  full outages where all computers at a site are down (perhaps due to a loss of power),
  and islands, where the site is cut off from the Internet but computers
    at the site can talk between themselves.
We next use Chiloe to determine how often islands occur.
We study islands in two systems with 6 \acp{VP} for 3 years
  and 13k \acp{VP} for 3 months.

\textbf{Trinocular:}
We first consider three years of Trinocular data (described in \autoref{sec:data_sources}),
  from 2017-04-01 to 2020-04-01.
We run Chiloe across each VP for this period.

\Cref{tab:island_summary} shows the number of islands per VP
  over this period.
Over the 3 years, all six \acp{VP} see from 1 to 5 islands.
In addition, we see that islands do not always cause the \emph{entire} Internet
  to be unreachable,
  and there are a number of cases where from 20\% to 50\% of the Internet is inaccessible.
We believe these cases represent brief islands, since
  islands shorter than an 11~minute complete scan will only be partially observed.
We find 12 in the 20\% to 50\% range,
  all are short, and 4 are less than 11 minutes
  (see \autoref{sec:islands_apendix} for details).

\textbf{RIPE Atlas:}
For broader coverage we next consider RIPE Atlas'
  13k \acp{VP} for all of 2021q3~\cite{ripe_ping}.
While Atlas does not scan the whole Internet,
  they do scan most root DNS servers every 240\,s.
Chiloe would like to observe the whole Internet, and
  while Trinocular scans 5M /24s,
  it does so with only 6 VPs.
To use RIPE Atlas' 10k VPs,
  we approximate a full scan
  with probes to 12 of the DNS root server systems (G-Root was unavailable in 2021q3).
Although far fewer than 5M networks,
  these targets provide a very sparse sample
  of usually independent destinations since each is independently operated.
Thus we have complementary datasets
  with sparse VPs and dense probing, and
  many VPs but sparse probing.
In other words, to get many VP locations
  we relax our conceptual definition by decreasing our target list.

\begin{figure*}
\begin{minipage}[b]{.25\textwidth}
	\begin{center}
            \includegraphics[trim=25 0 25 0,clip,width=0.99\textwidth]{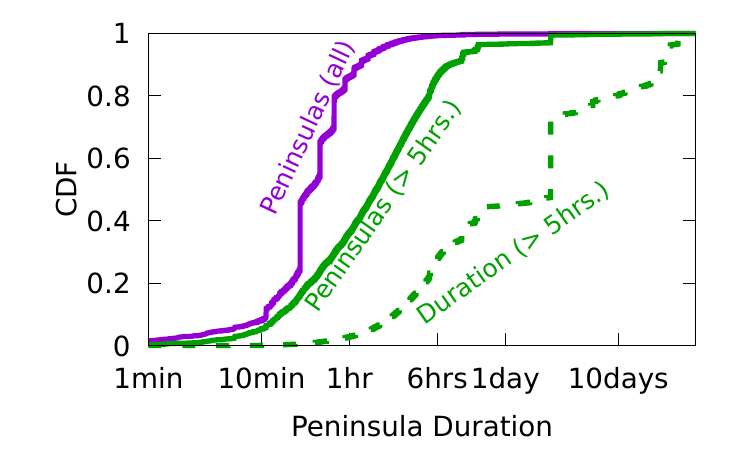}
	\end{center}
	\caption{Cumulative peninsulas and peninsula duration. Dataset A30, 2017q4.}
          \label{fig:a30_partial_outages_duration_cdf}
\end{minipage}
\hspace{-2mm}
\begin{minipage}[b]{.75\linewidth}
  \subfloat [Number of islands]{
    \includegraphics[trim=25 0 25 0,clip,width=0.33\linewidth]{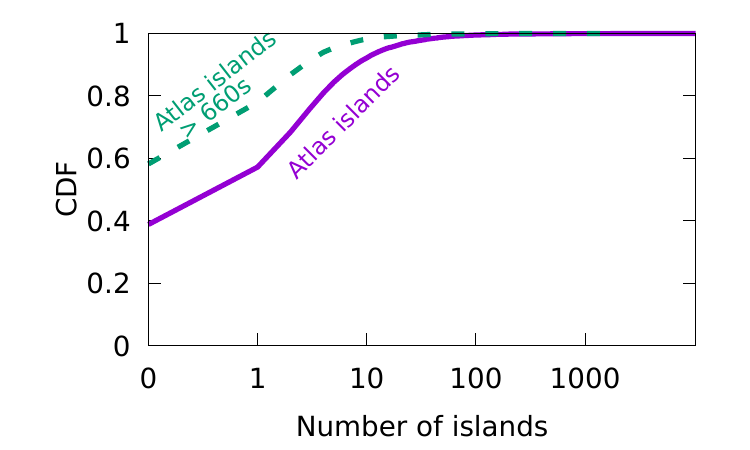}
    \label{fig:islands_per_node}
  }
  \subfloat[Duration of islands]{
    \includegraphics[trim=25 0 25 0,clip,width=0.33\textwidth]{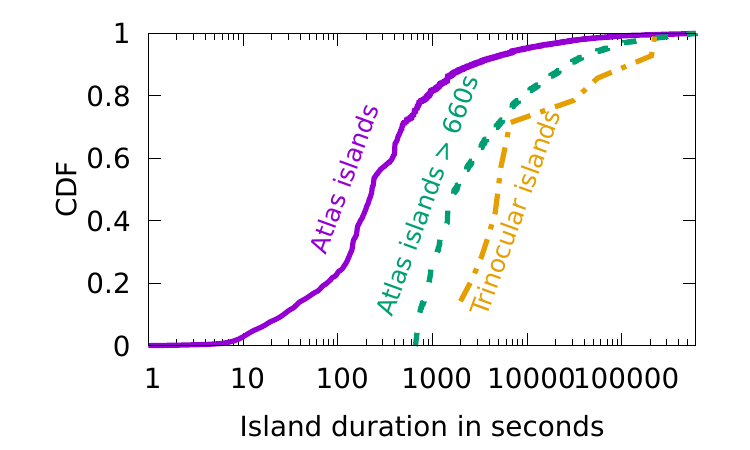}
    \label{fig:island_duration}
  }
  \subfloat[Size of islands]{
    \includegraphics[trim=25 0 25 0,clip,width=0.33\textwidth]{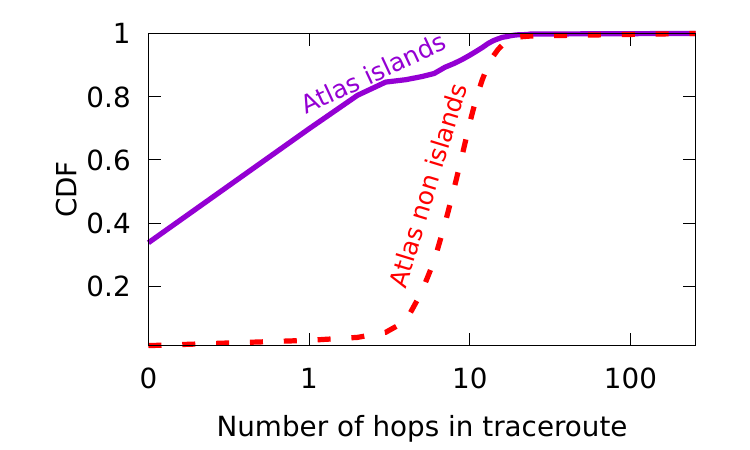}
    \label{fig:island_path_length}
  }
    \captionsetup{type=figure}
    \captionsetup{width=0.9\textwidth}
    \captionof{figure}{CDF of islands detected by Chiloe for data from Trinocular (3 years, Datasets A28-A39) and %
  Atlas (2021q3).}
        \label{fig:a30all_peninsulas_fig7}
  \end{minipage}
\end{figure*}

\autoref{fig:islands_per_node} shows the CDF of the number of islands detected
per RIPE Atlas \ac{VP} during 2021q3.
During this period, 55\% of \acp{VP} observed one or no islands (solid line).
To compare to Trinocular, we consider events longer than 660\,s with the dashed line.
In the figure,
  60\% of \acp{VP} saw no islands, 19\% see one, and the remainder see more.
The annualized island rate of just the most stable \acp{VP} (those that see 2 or less islands)
  is 1.75 islands per year (a lower bound, since we exclude less stable \acp{VP}),
  compared to 1.28 for Trinocular (\Cref{tab:island_summary}).
We see islands are more common in Atlas, perhaps because it includes
  many VPs at home.

We conclude that islands \emph{do} happen,
  but they are rare,
    and at irregular times.
This finding is consistent with importance of the Internet
  at the locations where we run VPs.

\subsection{How Long Do Islands Last?}
\label{sec:islands_duration}

Islands can occur starting from brief connectivity losses to
  long standing policy changes.
We next compare island duration measured across Trinocular and Atlas.

We compare the distributions of island durations observed from
  RIPE Atlas (the left line) and Trinocular (right) in
  \autoref{fig:island_duration}.
Since Atlas' frequent polling means it detects islands lasting seconds,
  while Trinocular sees only islands of 660\,s or longer,
  we split out Atlas events lasting at least 660\,s
  (middle line).
All measurements follow a similar S-shaped curve,
  but for Trinocular, the curve is truncated at 660\,s.
With only 6 VPs, Trinocular sees far fewer events (23 in 3 years compared to 235k in a yearly quarter with Atlas),
  so the Trinocular data is quantized.
In both cases, about 70\% of islands are between 1000 and 6000\,s.
This graph shows that Trinocular's curve is similar in shape to Atlas-660\,s,
  but about $2\times$ longer.
All Trinocular observers are in datacenters, while Atlas devices are at homes,
  so this difference may indicate that datacenter islands are rarer, but harder to resolve.

\subsection{What Sizes Are Islands?}
	\label{sec:islands_sizes}

In \autoref{sec:internet_landscape} we described different sizes of islands
  starting from as small as an address island,
  as opposed to LAN- or AS-sized islands,
  to country-sized islands potentially capable of partitioning the Internet.
Here, we evaluate the size of islands
  by counting the number of hops in a traceroute
  sent towards a target outside the island
  before the traceroute fails.

We use traceroutes from RIPE Atlas VPs sent to 12 root DNS servers
  for 2021q3 \cite{ripe_traceroute}.
In \autoref{fig:island_path_length} in green the distribution of the number of hops when traceroute reach their target.
In purple, we plot the distribution of the number of hops of traceroutes that failed to reach the target
for VPs in islands detected in \autoref{sec:how_common_are_islands}.

We find that most islands are small, 70\% show one hop or none (address islands).
We consider very large islands (10 or more hops) as false positives.

\section{Applying These Tools}

Given partial connectivity,
  we now apply our approach to
  Internet  sovereignty, partitioning,
  and DNSmon sensitivity.

\subsection{Policy Applications of the Definition}
	\label{sec:other_applications}

We next examine how a clear definition
  of the Internet's core can inform policy tussles~\cite{Clark02a}.
Our hope is that our conceptual definition can make
  sometimes amorphous concepts like ``Internet fragmentation''
  more concrete,
  and an operational definition can quantify impacts
  and identify thresholds.

\textbf{Secession and Sovereignty:}
The U.S.~\cite{cybersecurity_act_2010}, China~\cite{Anonymous12a,Anonymous14a},
and Russia~\cite{russian_internet} have all proposed unplugging from
the Internet.
Egypt did in 2011~\cite{Cowie11a},
  and several countries have during exams~\cite{Gibbs16a,Dhaka18a,Henley18a,Economist18a}.
When the Internet partitions,
  which part is still ``the Internet's core''?
Departure of a ISP or small country do not change the Internet's core much,
  but what if a large country, or group of countries, leave together?

Our definition resolves this question, defining the Internet's core
  from reachability of the majority of the active, public IP addresses
  (\autoref{sec:definition}).
Requiring a majority uniquely provides an unambiguous,
  externally evaluable test for the Internet's core
  that allows one possible answer (the partition with more than 50\%).
In \autoref{sec:internet_partition} we discuss the corollary:
  the internet can end, turning into multiple partitions,
  if none retain a majority.
(A plurality is insufficient.)

\textbf{Sanction:}
An opposite of secession is expulsion.
Economic sanctions are one method of asserting international influence,
  and events such as the 2022 war in Ukrainian prompted
  several large ISPs to discontinue service to Russia~\cite{Reuters22a}.
De-peering does not affect reachability for ISPs that purchase transit,
  but Tier-1 ISPs that de-peer create peninsulas for their users.
As described below in \autoref{sec:internet_partition},
  \emph{no single country can eject another by de-peering with it}.
However, a coalition of multiple countries could
  de-peer and eject a country from the Internet's core
  if they, together, control
  more than half of the address space.

\textbf{Repurposing Addresses:}
Given full allocation of IPv4,
  multiple parties proposed re-purposing currently allocated or reserved IPv4 space,
  such 0/8 (``this'' network), 127/8 (loopback), and 240/4 (reserved)~\cite{Fuller08a}.
New use of these long-reserved addresses is challenged
  by assumptions in widely-deployed, difficult to change, existing software
  and hardware.
Our definition demonstrates
  that an RFC re-assigning this space for public traffic
  cannot make it a truly effective part of the Internet core until
  implementations used by a majority of active addresses
  can route to it.

\textbf{IPv4 Squat Space:}
IP squatting is when an organization
  requiring private address space beyond RFC1918
  takes over allocated but currently unrouted IPv4 space~\cite{Aronson15a}.
Several IPv4 /8s allocated to the U.S.~DoD have been used this way~\cite{Richter16c}
  (they were only publicly routed in 2021~\cite{Timberg21a}).
By our definition, such space is not part of the Internet's core without
  public routes,
  and if more than half of the Internet is squatting on it,
  reclamation may be challenging.

\textbf{The IPv4/v6 Transition:}
We have defined two Internet cores: IPv4 and IPv6.
Our definition can determine when one supersedes the other.
The networks will be on par when
  more than half of all IPv4 hosts are dual-homed.
After that point, IPv6 will supersede IPv4 when
  a majority of hosts on IPv6 can no longer reach IPv4.
Current limits on IPv6 measurement mean evaluation
  here is future work.
IPv6 shows the strength and limits of our definition:
  since IPv6 is already economically important,
  our definition seems irrelevant.
However, it may provide sharp boundary
  that makes the maturity of IPv6 definitive,
  helping motivate late-movers.

\subsection{Can the Internet's Core Partition?}
	\label{sec:internet_partition}

In \autoref{sec:other_applications}
  we discussed secession and expulsion qualitatively.
Threats to secede or sanction have been by countries or groups of countries.
If a country were to exert control over their allocated addresses this would result
 in a country level island or peninsula.
We next use our reachability definition of more than 50\%
  to quantify control of the IP address space.
Our question: Does any
  country or group have enough addresses to secede and claim to be
  ``the Internet's core'' with a majority of addresses.

\begin{table*}
\begin{minipage}[b]{.33\linewidth}
    \footnotesize
    \resizebox{1\textwidth}{!}{
    \begin{tabular}{l r r r r r r}
      & \multicolumn{4}{c}{\textbf{IPv4 Addresses}} &
      \multicolumn{2}{c}{\textbf{IPv6 Addresses}} \\
      \textbf{RIR}       &
      \multicolumn{2}{c}{\textbf{Active}}  &
      \multicolumn{2}{c}{\textbf{Allocated}}  &
      \multicolumn{2}{c}{\textbf{Allocated}} \\
      \midrule
      AFRINIC   & 15M    &  2\%   & 121M   & 3.3\%      & 9,661   & 3\%       \\
      APNIC     & 223M   & \cellcolor[HTML]{99ee77}33\%   & 892M  & 24.0\%      & 88,614  & 27.8\%    \\
      \rowcolor[HTML]{DCDCDC}
      \hspace{1mm} \emph{China}   & 112M & 17\% &  345M   &    9.3\%     & 54,849  & \cellcolor[HTML]{FFF9C4}17.2\%    \\
      ARIN      & 150M   & 22\%   & 1,673M & \cellcolor[HTML]{99ee77}45.2\%      & 56,172  & 17.6\%    \\
      \rowcolor[HTML]{DCDCDC}
      \hspace{1mm} \emph{U.S.}    & 140M & \cellcolor[HTML]{FFF9C4}21\% & 1,617M  & \cellcolor[HTML]{FFF9C4}43.7\%        & 55,026  & \cellcolor[HTML]{FFF9C4}17.3\%  \\
      LACNIC    & 82M  & 12\%     & 191M  & 5.2\%      & 15,298  & 4.8\%     \\
      RIPE NCC  & 206M & 30\%     & 826M  & 22.3\%      & 148,881 & \cellcolor[HTML]{99ee77}46.7\%    \\
      \rowcolor[HTML]{DCDCDC}
      \hspace{1mm} \emph{Germany} & 40M & 6\% & 124M  &    3.3\%     & 22,075  & 6.9\%     \\
      \midrule
      Total &  676M &  100\% & 3,703M & 100\%      & 318,626 & 100\%     \\
    \end{tabular}}
    \vspace{5mm}
    \caption{RIR IPv4 hosts and IPv6 /32 allocation
    \cite{iana_v4, iana_v6}}
    \label{tab:rir_allocation}
\end{minipage}
\begin{minipage}[b]{.33\linewidth}
    \centering
    \footnotesize
    \resizebox{1\textwidth}{!}{
    \includegraphics[trim=0 0 0 5,clip,width=1\textwidth]{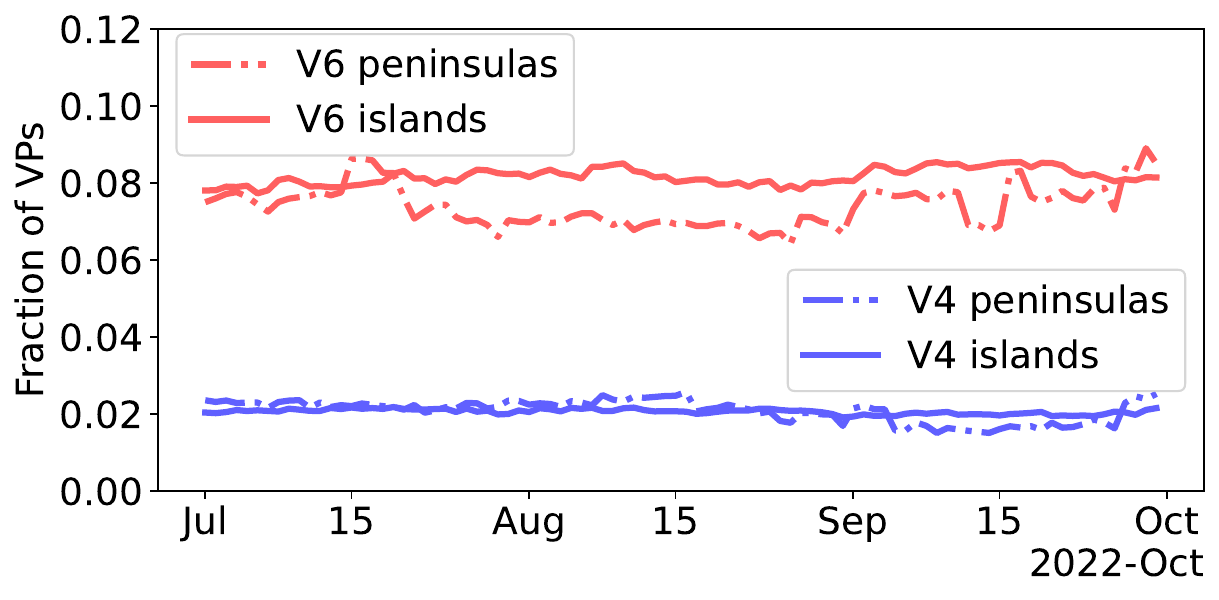}}
    \captionsetup{type=figure}
    \captionsetup{width=0.9\textwidth}
    \captionof{figure}{Fraction of VPs observing islands and peninsulas for IPv4 and IPv6 during 2022q3}
  \label{fig:a49_partial_outages}
\end{minipage}
\begin{minipage}[b]{.33\linewidth}
    \centering
    	\footnotesize
        \resizebox{1\textwidth}{!}{
        \includegraphics[trim=0 0 0 5,clip,width=1\textwidth]{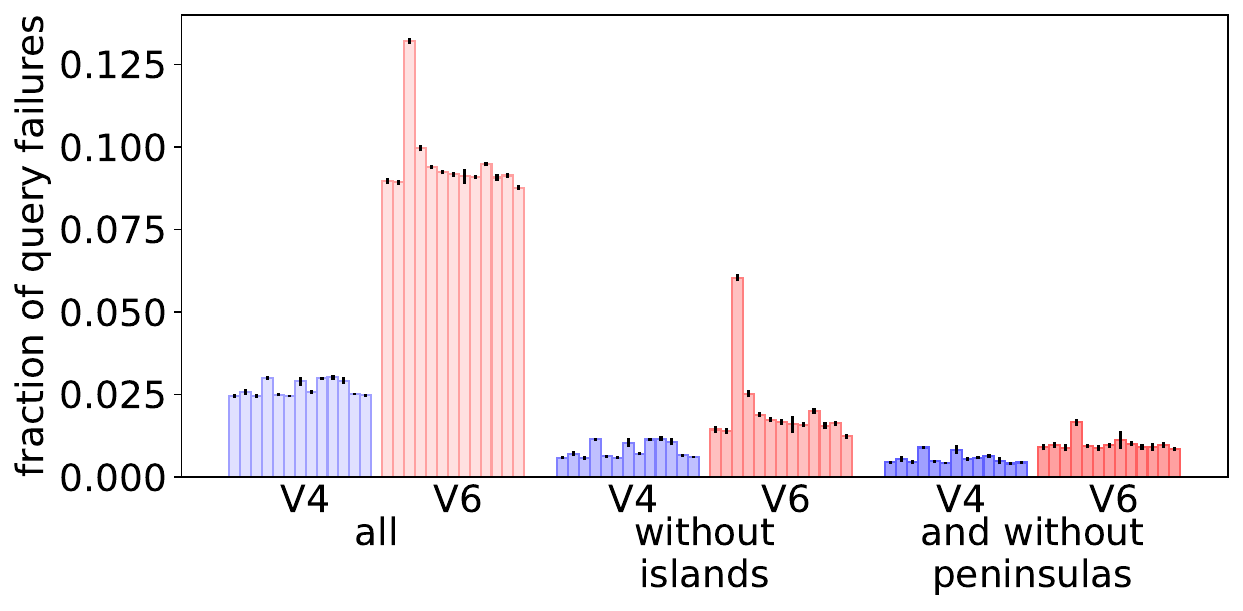}}
    \captionsetup{type=figure}
    \captionof{figure}{Atlas queries from all available VPs to 13
    Root Servers for IPv4 and IPv6 on 2022-07-23.}
    \label{fig:atlas_revisited}
\end{minipage}
\end{table*}

To evaluate the power of any country or RIR to control the Internet core,
\autoref{tab:rir_allocation} reports the number of active IPv4
  addresses as determined by Internet censuses~\cite{Heidemann08c}
  for each
  Regional Internet Registry (RIR) and selected countries.
Although we define the Internet by active addresses,
  we cannot current measure active IPv6 addresses,
  so we also provide allocated addresses for both v4 and v6~\cite{iana_v4, iana_v6}.
IPv4 is fully allocated, except for special purpose addresses:
  loopback (127/8), local and private space (0/8, 10/8, etc.~\cite{Rekhter96a}),
  multicast, and reserved Class E addresses.

\autoref{tab:rir_allocation} shows that no individual RIR or country can secede and take the Internet's core,
  because none controls the majority of IPv4 addresses.
ARIN has the largest share with 1673M allocated (45.2\%).
Of countries, U.S. has the largest share of allocated IPv4 (1617M, 43.7\%).
Active addresses are more evenly distributed
  with APNIC (223M, 33\%) and the U.S.~(40M, 21\%) the largest RIR and country.

This claim also applies to IPv6, where  no RIR or country surpasses a 50\% allocation.
RIPE (an RIR) is close with 46.7\%,
  and China and the U.S.~have large country allocations.
With most of IPv6 unallocated, these fractions may change.
Distribution
  of active IPv4 addresses is similar to allocated IPv6 addresses,
  suggesting IPv4 allocations are perhaps skewed by unused legacy addresses.

Our analysis demonstrates that \emph{no country can
  unilaterally claim to control the IPv4 Internet core,
  nor the currently allocated IPv6 core}---today's Internet is an international collaboration.

\subsection{Reexamining Outages Given Partial Reachability}
	\label{sec:local_outage_eval}
	\label{sec:representing_the_internet}

We next re-evaluate reports from existing outage detection systems,
  considering how to resolve conflicting information in light of
  our new algorithms.
We compare findings to external information in traceroutes from CAIDA Ark.

  \begin{figure}
  \centering
  \includegraphics[width=.75\columnwidth]{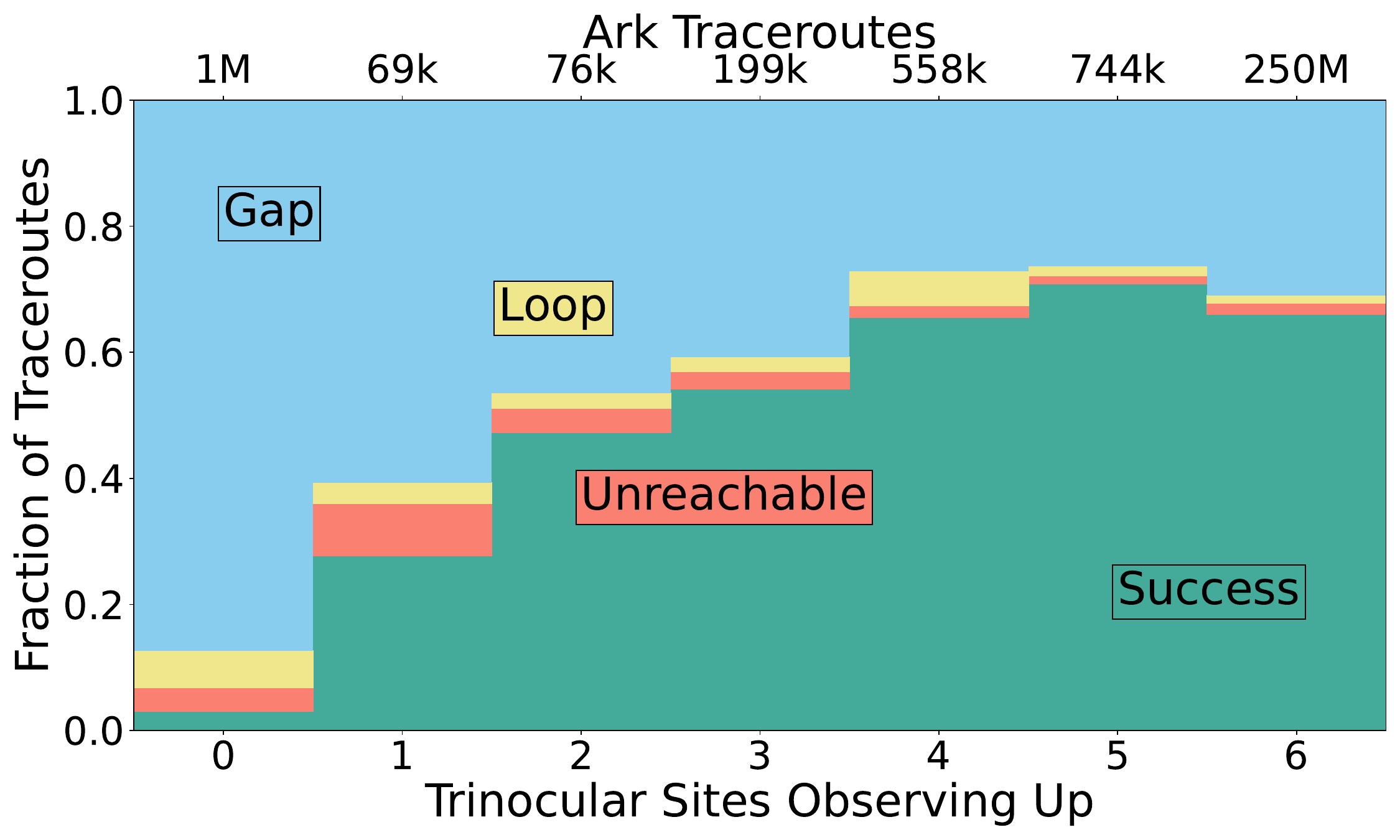}
        \caption{Ark traceroutes sent to targets under partial outages
        (2017-10-10 to -31). Dataset A30.}
  \label{fig:a30all_reach_fraction}
  \end{figure}

\autoref{fig:a30all_reach_fraction} compares Trinocular with
  21 days of Ark topology data, from 2017-10-10 to -31 from all 3 probing teams.
For each Trinocular outage we classify the Ark result as success
  or three types of failure: unreachable, loop, or gap.

\begin{table}
    \footnotesize
    \centering
	\begin{tabular}{c|c@{\hspace{0.7ex}}c@{\hspace{0.7ex}}c@{\hspace{0.7ex}}c@{\hspace{0.7ex}}c}
       & C      & J      & G      & E      & N      \\
	\hline
	W  & 0.017  & 0.031  & 0.019  & 0.035  & 0.020  \\
    C  &        & 0.077  & 0.143  & 0.067  & 0.049  \\
    J  &        &        & 0.044  & 0.036  & 0.046  \\
    G  &        &        &        & 0.050  & 0.100  \\
    E  &        &        &        &        & 0.058  \\
    \end{tabular}
    \caption{Similarities between sites relative to all six. Dataset: A30, 2017q4.}
    \label{tab:overall_correlation}
\end{table}

Trinocular's 6-site-up case suggests a working network,
  and we consider this case as typical.
However, we see that about 25\% of Ark traceroutes are ``gap'',
  where several hops fail to reply.
We also see about 2\% of traceroutes are unreachable
  (after we discard traceroutes to never reachable addresses).
Ark probes a random address in each block;
  many addresses are non-responsive,
  explaining these.

With 1 to 11 sites up, Trinocular is reporting disagreement.
We see that the number of Ark success cases (the green, lower portion of each bar)
  falls roughly linearly with the number of
  successful observers.
This consistency suggests that Trinocular and Ark are seeing similar behavior,
  and that there is partial reachability---these events
  with only partial Trinocular positive results
  are peninsulas.

We observe that 5 sites show the same results as all 6,
  so single-\ac{VP} failures likely represent problems local to that \ac{VP}.
This suggests that all-but-one is a good algorithm to determine true outages.

With only partial reachability, with 1 to 4 VPs (of 6),
  we see likely peninsulas.
These cases confirm that partial connectivity is common:
  while there are 1M traceroutes sent to outages where no \ac{VP} can see the target
  (the number of events is shown on the 0 bar),
  there are 1.6M traceroutes sent to partial outages
  (bars 1 to 5),
  and 850k traceroutes sent to definite peninsulas (bars 1 to 4).
This result is consistent with the convergence we see in
\autoref{fig:a30all_peninsulas_duration_oct_nov}.

\subsection{Improving DNSmon Sensitivity}
	\label{sec:dnsmon}

DNSmon~\cite{Amin15a}
  monitors the Root Server System~\cite{RootServers16a},
  built over the RIPE Atlas distributed platform~\cite{Ripe15c}
For years, DNSmon has often reported IPv6 loss rates of 4-10\%.
Since the DNS root is well provisioned and distributed,
  we expect minimal congestion or loss
  and find these values surprisingly high.

RIPE Atlas operators are aware of problems with some Atlas VPs.
Some support IPv6 on their LAN, but not to the global IPv6 Internet---such VPs
  are IPv6 islands.
They periodically tag these VPs and cull
  them from DNSmon.
However, we studied RIPE Atlas
  with our
  algorithms to detect islands and peninsulas.
Full details of our analysis are in our
    workshop paper~\cite{Saluja22a};
  but it builds on the concepts pioneered here (\autoref{sec:problem} and \autoref{sec:design}).
We also provide the first long-term data
  that shows these results persist for 4 months (\autoref{fig:a49_partial_outages}).

Each groups of bars in
  \autoref{fig:atlas_revisited} show query loss
  for each of the 13 root service identifiers,
  as observed
  from all available Atlas VPs (10,082 IPv4, and 5,173 IPv6)
  on 2022-07-23.
(We are similar to DNSmon, but it uses only about 100 well-connected ``anchors'',
  so our analysis is wider.)
The first two groups show loss rates for IPv4 (light blue, left most) and IPv6 (light red),
  showing IPv4 losses around 2\%, and IPv6 from 9 to 13\%.

We apply Chiloe to these VPs, detecting as islands those VPs that
  cannot see \emph{any} of the 13 root identifiers over 24~hours.
(This definition is stricter than regular Chiloe because these VPs attempt only 13 targets,
  and we apply it over a full day to consider only long-term trends.)
The middle two groups of bars show IPv4 and IPv6 loss rates
  after removing VPs that
  are islands.
Without island VPs,
  IPv4 loss rates drop to 0.005 from 0.01, and IPv6 to about 0.01 from 0.06.
We suggest this represents a more
  accurate view of how most people perceive the root queries.
Islands represent misconfigured VPs; they should not be used for measurement
  until they can route outside their LAN\@.

The third bar in each red cluster of IPv6 is an outlier:
  that root identifier shows 13\% IPv6 loss with all VPs,
  and 6\% loss after islands are removed.
This result is explained by
  persistent routing disputes between Cogent (the operator of C-Root) and Hurricane Electric~\cite{Miller09a}.
Omitting islands (the middle bars) makes this difference is much clearer.

We then apply Taitao to detect peninsulas.
Peninsulas suggest persistent routing problems
  deserving consideration by ISPs and root operators.
The darker, rightmost two groups show loss from VPs that are
  neither islands nor peninsulas,
  representing loss if routing problems were addressed.
This data confirms routing problems explain the difference
  for C-Root, which now shows IPv6 loss similar to other identifiers.

This example shows how \emph{understanding of partial reachability
  can improve the sensitivity of existing measurement systems}.
Filtering out islands makes it easy to identify persistent routing problems.
Further removing peninsulas
  leaves observations that are more sensitive to
  transient changes, perhaps from failure, DDoS attack,
  or temporary routing changes---\autoref{fig:atlas_revisited} shows
  that the raw data (left two groups)
  are $5\times$ or $9.7\times$ times larger than this remaining interesting ``signal'' (the right two groups).
Improved sensitivity also clarifies the need to improve
  IPv6 provisioning,
  since
  IPv6 loss is statistically higher than
  IPv4 loss (compare the right blue and red groups), even after correcting for known problems.

While application of our algorithms to this system is imperfect,
  we suggest that it is useful.
Atlas VPs do not ping the entire Internet,
  so our evaluation of islands over the 13 root identifiers
  is very rough.
While we suggest islands represent misconfiguration, peninsulas
  show actual, persistent connectivity problems
  (fortunately not harming users because of the redundancy with 13 separate services).
We have shared these results with several root operators
  and RIPE Atlas;
  two operators are using them in regular operation,
  showing their utility.

\section{Related Work}
	\label{sec:related}

A number of works have suggested definitions of the
    Internet~\cite{Cerf74a,Postel80b,nitrd,huawei2020}.
As discussed in \autoref{sec:problem}, %
  they distinguish the Internet from other networks of their time,
  but do not address today's network disputes and secession threats.

Previous work has looked into the problem of partial outages.
RON provides alternate-path routing around failures for a mesh of sites~\cite{andersen2001resilient}.
HUBBLE monitors in real-time reachability problems
  in which a working physical path exists.
LIFEGUARD, proposes a route failure
remediation system by generating BGP messages to reroute traffic through a
working path~\cite{katz2012lifeguard}.
While both solve the problem of partial outages,
  neither quantifies the amount, duration, or scope of partial
  outages in the Internet.

Prior work studied partial reachability, showing
  it is a common transient occurrence
  during routing convergence~\cite{bush2009internet}.
They reproduced partial connectivity with controlled experiments;
  we study it from Internet-wide vantage points.

Internet scanners have examined bias by location~\cite{Heidemann08c},
  more recently looking for policy-based filtering~\cite{wan2020origin}.
  We measure policies with our country specific algorithm, and
  we extend those ideas to defining the Internet.

Outage detection systems have encountered partial outages.
Thunderping recognizes the ``hosed'' state of partial replies as something that occurs,
  but leaves its study to future work~\cite{Schulman11a}.
Trinocular discards partial outages by
  reporting the target block ``up'' if any VP can reach
  it~\cite{quan2013trinocular}.
To the best of our knowledge, prior outage detection systems
  have not both explained and reported partial outages as part of the Internet,
  nor studied their extent.

We use the idea of majority to define the Internet in the face of secession.
That idea is fundamental in many algorithms for distributed consensus~\cite{Lamport82a,Lamport98a,Nakamoto09a},
  with applications for example to certificate authorities~\cite{birge2018bamboozling}.

Recent groups have studied the policy issues around Internet fragmentation~\cite{Drake16a,Drake22a}, but do not define it.
We hope our definition can fill that need.

\section{Conclusions}

This paper provided a new definition of the Internet's core.
We developed the algorithm Taitao, to find peninsulas of partial connectivity,
  and Chiloe, to find islands.
We showed that partial connectivity events are more common than simple outages,
  and suggest they help clarify questions around Internet sovereignty and evolution.

\begin{acks}

The authors would like to thank John Wroclawski,
  Wes Hardaker, Ramakrishna Padmanabhan,
  Ramesh Govindan,
  Eddie Kohler,
  and the Internet Architecture Board for their input on
  on an early version of this paper.

The work is
  supported in part by
   the National Science Foundation, CISE
  Directorate, award CNS-2007106 %
  and NSF-2028279. %
The U.S.~Government is authorized to reproduce and distribute
reprints for Governmental purposes notwithstanding any copyright
notation thereon.
\end{acks}

\label{page:last_body}

\bibliographystyle{ACM-Reference-Format}

\appendix

\section{Research Ethics}
	\label{sec:research_ethics}

Our work poses no ethical concerns for several reasons.

First, we collect no additional data, but instead
  reanalyze data from several existing sources
  listed in \autoref{sec:data_sources_list}.
Our work therefore poses no additional risk in data collection.

Our analysis poses no risk to individuals
  because our subject is network topology and connectivity.
There is a slight risk to individuals in that we
  examine responsiveness of individual IP addresses.
With external information, IP addresses can sometimes be traced to individuals,
  particularly when combined with external data sources like DHCP logs.
We avoid this risk in three ways.
First, we do not have DHCP logs for any networks
  (and in fact, most are unavailable outside of specific ISPs).
Second, we commit, as research policy, to not combine
  IP addresses with external data sources
  that might de-anonymize them to individuals.
Finally, except for analysis of specific cases as part of validation,
  all of our analysis is done in bulk over the whole dataset.

We do observe data about organizations such as ISPs,
  and about the geolocation of blocks of IP addresses.
Because we do not map IP addresses to individuals,
  this analysis poses no individual privacy risk.

Finally, we suggest that while our work poses minimal privacy risks
  to individuals,
  to also provides substantial benefit to the community and to individuals.
For reasons given in the introduction
  it is important to improve network reliability and understand
  now networks fail.
Our work contributes to that goal.

Our work was reviewed by the
  Institutional Review Board at our university
  and because it poses no risk to individual privacy,
  it was identified as non-human subjects research
  (USC IRB IIR00001648).

\section{Data Sources Used Here }
	\label{sec:data_sources_list}

\begin{table*}
  \resizebox{\textwidth}{!}{%
	\begin{tabular}{llllp{6.5cm}}
	\textbf{Dataset Name} & \textbf{Source} & \textbf{Start Date} & \textbf{Duration} & \textbf{Where Used} \\
	\hline
	\rowcolor[HTML]{DCDCDC}
	internet\_outage\_adaptive\_a28w-20170403 & Trinocular~\cite{trinoculardatasets} & 2017-04-03 & 90 days &  \\
	\rowcolor[HTML]{DCDCDC}
	\quad Polish peninsula subset & & 2017-06-03 & 12 hours & \autoref{sec:island}, \autoref{sec:polish_peninsula_validation} \\
	internet\_outage\_adaptive\_a28c-20170403 & Trinocular & 2017-04-03 & 90 days & \\
	\quad Polish peninsula subset & & 2017-06-03 & 12 hours & \autoref{sec:polish_peninsula_validation}\\
	\rowcolor[HTML]{DCDCDC}
	internet\_outage\_adaptive\_a28j-20170403 & Trinocular & 2017-04-03 & 90 days & \\
	\rowcolor[HTML]{DCDCDC}
	\quad Polish peninsula subset & & 2017-06-03 & 12 hours & \autoref{sec:polish_peninsula_validation}\\
	internet\_outage\_adaptive\_a28g-20170403 & Trinocular & 2017-04-03 & 90 days & \\
	\quad Polish peninsula subset & & 2017-06-03 & 12 hours & \autoref{sec:polish_peninsula_validation}\\
	\rowcolor[HTML]{DCDCDC}
	internet\_outage\_adaptive\_a28e-20170403 & Trinocular & 2017-04-03 & 90 days & \\
	\rowcolor[HTML]{DCDCDC}
	\quad Polish peninsula subset & & 2017-06-03 & 12 hours & \autoref{sec:island}, \autoref{sec:polish_peninsula_validation} \\
	internet\_outage\_adaptive\_a28n-20170403 & Trinocular & 2017-04-03 & 90 days & \\
	\quad Polish peninsula subset & & 2017-06-03 & 12 hours & \autoref{sec:island}, \autoref{sec:polish_peninsula_validation} \\
	\rowcolor[HTML]{DCDCDC}
	internet\_outage\_adaptive\_a28all-20170403 & Trinocular & 2017-04-03 & 89 days & \autoref{sec:chiloe_validation},
  											   \autoref{sec:how_common_are_islands},
    											   \autoref{sec:islands_duration},
  											   \autoref{sec:islands_apendix} \\
	internet\_outage\_adaptive\_a29all-20170702 & Trinocular & 2017-07-02 & 94 days  & \autoref{sec:island},
											   \autoref{sec:chiloe_validation},
  											   \autoref{sec:how_common_are_islands},
    											   \autoref{sec:islands_duration},
  											   \autoref{sec:islands_apendix} \\
	\rowcolor[HTML]{DCDCDC}
	internet\_outage\_adaptive\_a30w-20171006 & Trinocular & 2017-10-06 & 85 days & \\
	\rowcolor[HTML]{DCDCDC}
	\quad Site E Island & & 2017-10-23 & 36 hours & \autoref{sec:peninsula_definition}, \autoref{sec:polish_peninsula_validation} \\
	internet\_outage\_adaptive\_a30c-20171006 & Trinocular & 2017-10-06 & 85 days & \\
	\quad Site E Island & & 2017-10-23 & 36 hours & \autoref{sec:polish_peninsula_validation} \\
	\rowcolor[HTML]{DCDCDC}
	internet\_outage\_adaptive\_a30j-20171006 & Trinocular & 2017-10-06 & 85 days & \\
	\rowcolor[HTML]{DCDCDC}
	\quad Site E Island & & 2017-10-23 & 36 hours & \autoref{sec:polish_peninsula_validation} \\
	internet\_outage\_adaptive\_a30g-20171006 & Trinocular & 2017-10-06 & 85 days & \\
	\quad Site E Island & & 2017-10-23 & 36 hours & \autoref{sec:polish_peninsula_validation} \\
	\rowcolor[HTML]{DCDCDC}
	internet\_outage\_adaptive\_a30e-20171006 & Trinocular & 2017-10-06 & 85 days & \\
	\rowcolor[HTML]{DCDCDC}
	\quad Site E Island & & 2017-10-23 & 36 hours & \autoref{sec:peninsula_definition}, \autoref{sec:polish_peninsula_validation} \\
	internet\_outage\_adaptive\_a30n-20171006 & Trinocular & 2017-10-06 & 85 days & \\
	\quad Site E Island & & 2017-10-23 & 36 hours & \autoref{sec:peninsula_definition}, \autoref{sec:polish_peninsula_validation} \\
	\rowcolor[HTML]{DCDCDC}
	internet\_outage\_adaptive\_a30all-20171006 & Trinocular & 2017-10-06 & 85 days  & \autoref{sec:chiloe_validation},
  											   \autoref{sec:how_common_are_islands},
    											   \autoref{sec:islands_duration},
    											   \autoref{sec:site_correlation},
  											   \autoref{sec:islands_apendix} \\
	\rowcolor[HTML]{DCDCDC}
	\quad   Oct. Nov. subset & & 2017-10-06 & 40 days  &
	\autoref{sec:country_validation},
											   \autoref{sec:peninsula_duration},
											   \autoref{sec:peninsula_size}\\
	\rowcolor[HTML]{DCDCDC}
	\quad	Oct. subset & & 2017-10-10 & 21 days  & \autoref{sec:taitao_validation},
  											   \autoref{sec:representing_the_internet} \\
	internet\_outage\_adaptive\_a31all-20180101 & Trinocular & 2018-01-01 & 90 days  & \autoref{sec:chiloe_validation},
  											   \autoref{sec:how_common_are_islands},
    											   \autoref{sec:islands_duration},
  											   \autoref{sec:islands_apendix} \\
	\rowcolor[HTML]{DCDCDC}
	internet\_outage\_adaptive\_a32all-20180401 & Trinocular & 2018-04-01 & 90 days  & \autoref{sec:chiloe_validation},
  											   \autoref{sec:how_common_are_islands},
    											   \autoref{sec:islands_duration},
  											   \autoref{sec:islands_apendix}\\
	internet\_outage\_adaptive\_a33all-20180701 & Trinocular & 2018-07-01 & 90 days  & \autoref{sec:chiloe_validation},
  											   \autoref{sec:how_common_are_islands},
    											   \autoref{sec:islands_duration},
  											   \autoref{sec:islands_apendix}\\
	\rowcolor[HTML]{DCDCDC}
	internet\_outage\_adaptive\_a34all-20181001 & Trinocular & 2018-10-01 & 90 days  & \autoref{sec:chiloe_validation},
  											   \autoref{sec:how_common_are_islands},
    											   \autoref{sec:islands_duration},
											   \autoref{sec:additional_confirmation},
  											   \autoref{sec:islands_apendix}\\
	internet\_outage\_adaptive\_a35all-20190101 & Trinocular & 2019-01-01 & 90 days  & \autoref{sec:chiloe_validation},
  											   \autoref{sec:how_common_are_islands},
    											   \autoref{sec:islands_duration},
  											   \autoref{sec:islands_apendix}\\
	\rowcolor[HTML]{DCDCDC}
	internet\_outage\_adaptive\_a36all-20190401 & Trinocular & 2019-01-01 & 90 days  & \autoref{sec:chiloe_validation},
  											   \autoref{sec:how_common_are_islands},
    											   \autoref{sec:islands_duration},
  											   \autoref{sec:islands_apendix}\\
	internet\_outage\_adaptive\_a37all-20190701 & Trinocular & 2019-01-01 & 90 days  & \autoref{sec:chiloe_validation},
  											   \autoref{sec:how_common_are_islands},
    											   \autoref{sec:islands_duration},
  											   \autoref{sec:islands_apendix}\\
	\rowcolor[HTML]{DCDCDC}
	internet\_outage\_adaptive\_a38all-20191001 & Trinocular & 2019-01-01 & 90 days  & \autoref{sec:chiloe_validation},
  											   \autoref{sec:how_common_are_islands},
    											   \autoref{sec:islands_duration},
  											   \autoref{sec:islands_apendix}\\
	internet\_outage\_adaptive\_a39all-20200101 & Trinocular & 2020-01-01 & 90 days  & \autoref{sec:chiloe_validation},
  											   \autoref{sec:how_common_are_islands},
    											   \autoref{sec:islands_duration},
  											   \autoref{sec:islands_apendix}\\
	\rowcolor[HTML]{DCDCDC}
	internet\_outage\_adaptive\_a41all-20200701 & Trinocular & 2020-07-01 & 90 days  & \autoref{sec:peninsula_locations} \\
	\hline
	prefix-probing & Ark~\cite{CAIDA07b} \\
	\quad Oct. 2017 subset & & 2017-10-10 & 21 days & \autoref{sec:taitao_validation}, \autoref{sec:representing_the_internet} \\
	\quad 2020q3 subset & & 2020-07-01 & 90 days & \autoref{sec:peninsula_locations}\\
	\rowcolor[HTML]{DCDCDC}
	probe-data     & Ark & & & \\
	\rowcolor[HTML]{DCDCDC}
	\quad Oct 2017 subset     & & 2017-10-10 & 21 days & \autoref{sec:taitao_validation}, \autoref{sec:representing_the_internet} \\
	\rowcolor[HTML]{DCDCDC}
	\quad 2020q3 subset     & & 2020-07-01 & 90 days & \autoref{sec:peninsula_locations}\\
	\hline
	routeviews.org/bgpdata & Routeviews~\cite{routeviews} & 2017-10-06 & 40 days
    &
      \autoref{sec:country_validation},
  								      \autoref{sec:polish_peninsula_validation} \\
	\hline
	\rowcolor[HTML]{DCDCDC}
	Atlas Recurring Root Pings (id: 1001 to 1016) & Atlas~\cite{ripe_ping} & 2021-07-01 & 90 days & \autoref{sec:peninsula_frequency},
													\autoref{sec:islands_duration} \\
	\hline
	nro-extended-stats & NRO~\cite{iana_v4,iana_v6} & 1984 & 41 years & \autoref{sec:internet_partition} \\

	\end{tabular}
  }
   \caption{All datasets used in this paper.}
   \label{tab:datasets}
\end{table*}

\autoref{tab:datasets} provides a full list of datasets used in this paper
  and where they may be obtained.

\section{Validation of the Polish Peninsula}
	\label{sec:polish_peninsula_validation}

On 2017-10-23, for a period of 3 hours starting at 22:02Z,
  five Polish \acp{AS}
  had 1716 blocks that were unreachable from five \acp{VP}
  while the same blocks remained reachable from a sixth \ac{VP}.

\begin{figure}
  \includegraphics[width=1\columnwidth]{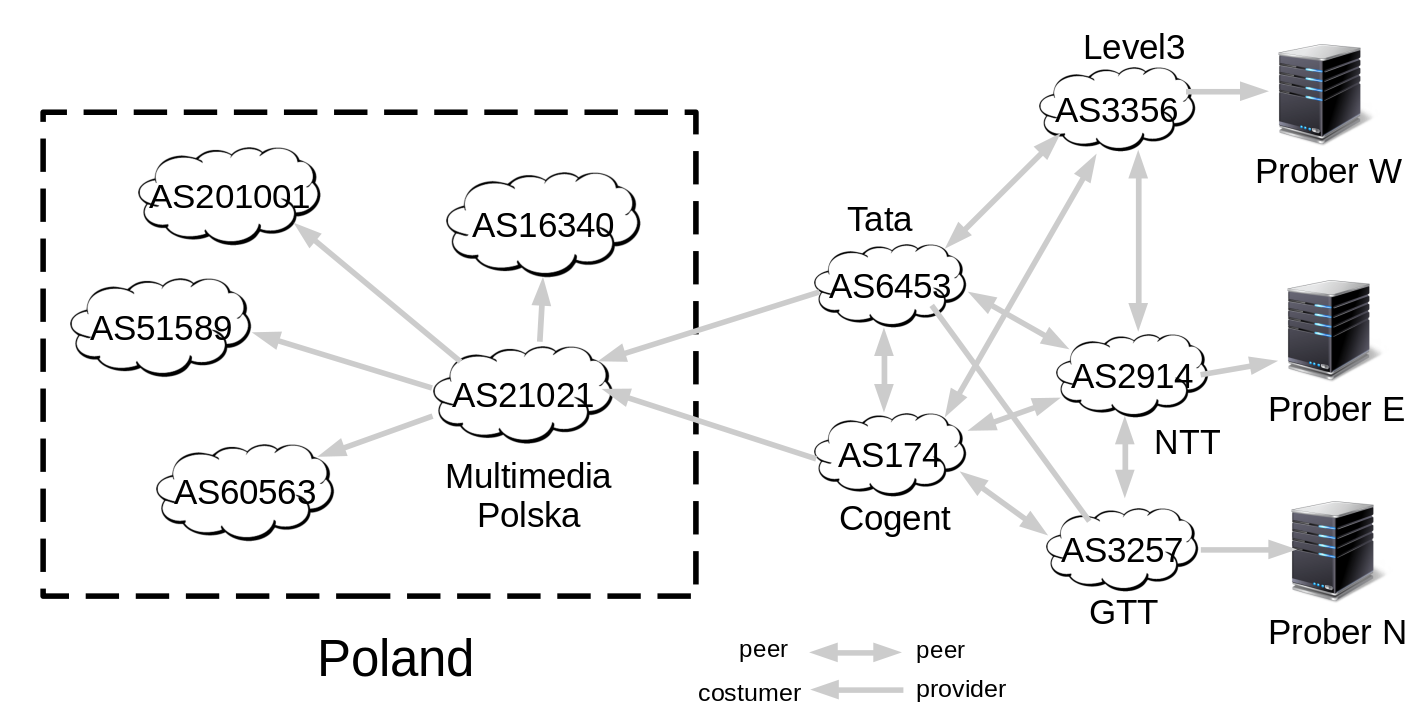}
  \caption{AS level topology during the Polish peninsula.}
  \label{fig:polish_peninsula}
\end{figure}

\autoref{fig:polish_peninsula} shows the AS-level relationships
  at the time of the peninsula.
Multimedia Polska (AS21021, or \emph{MP}) provides service to the other 4 ISPs.
MP has two Tier-1 providers:
  Cogent (AS174) and Tata (AS6453).
Before the peninsula, our VPs see MP
  through Cogent.

\begin{figure}
  \includegraphics[width=0.9\columnwidth]{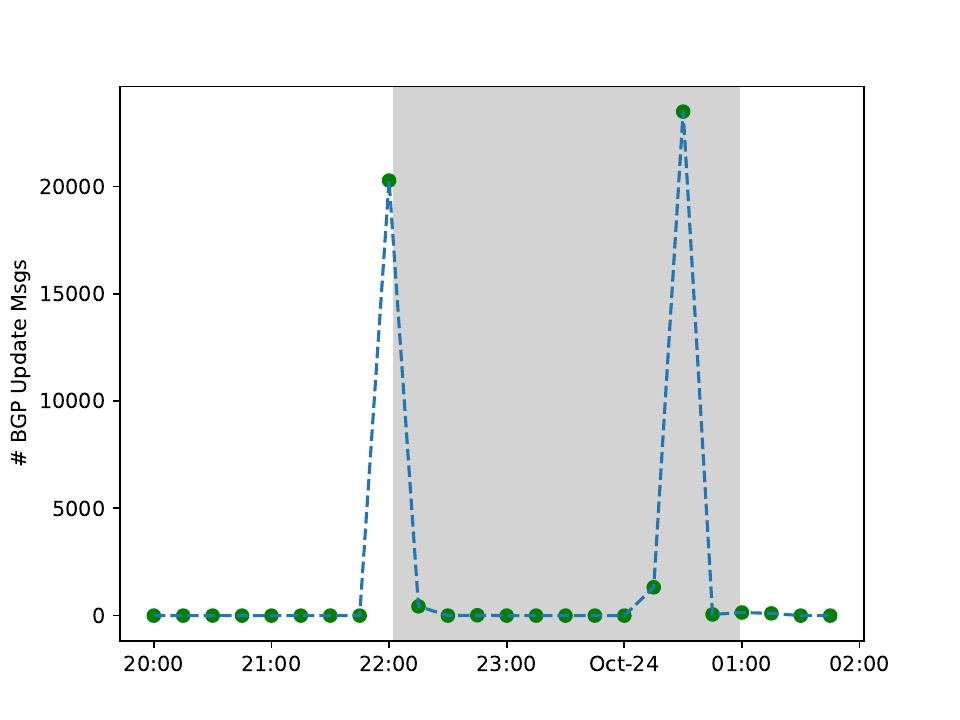}
	\caption{ BGP update messages sent for affected Polish blocks starting
    2017-10-23t20:00Z. Data source: RouteViews. }
  \label{fig:rv_time_plot}
\end{figure}

At event start, we observe many BGP updates (20,275) announcing
and withdrawing routes to the affected blocks(see~\autoref{fig:rv_time_plot}).
These updates correspond to Tata announcing MP's prefixes.
Perhaps MP changed its peering
  to prefer Tata over Cogent,
  or the MP-Cogent link failed.

Initially, traffic from most VPs continued through Cogent
  and was lost; it did not shift to Tata.
One \ac{VP} (W) could reach MP
  through  Tata for the entire event,
  proving MP was connected.
After 3 hours, we see another burst of BGP updates (23,487 this time),
  making MP reachable again from all VPs.

\begin{figure}
  \includegraphics[width=1\columnwidth]{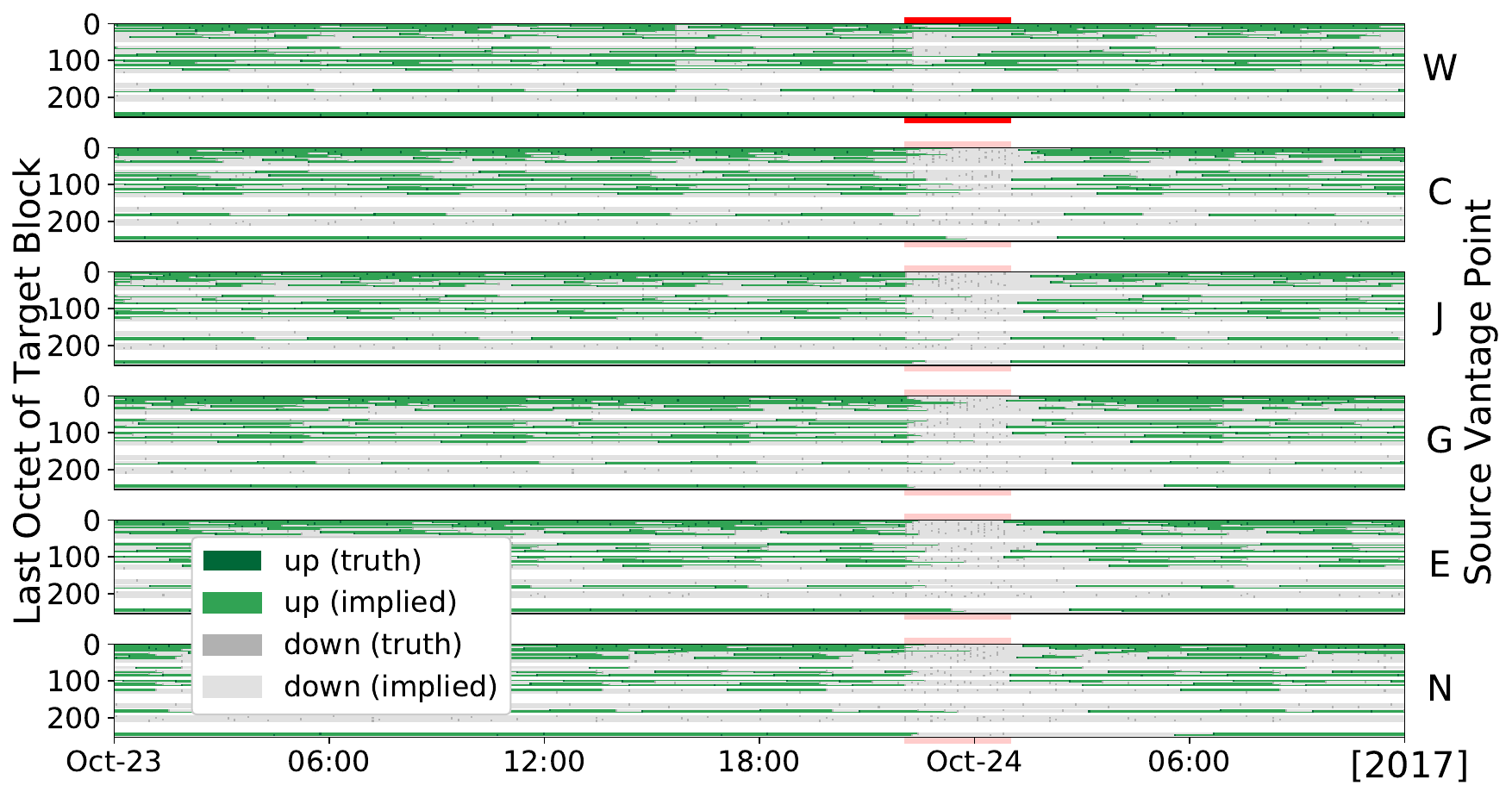}
  \caption{A block
            (80.245.176.0/24)
      showing a 3-hour peninsula accessible only from \ac{VP} W (top bar)
      and not from the other five \acp{VP}.  Dataset: A30.}
  \label{fig:a30all_raw_50f5b000_6sites}
\end{figure}

In \autoref{fig:a30all_raw_50f5b000_6sites} we provide data from our 6 external VPs,
where W is uniquely capable of reaching the target block, thus living in the
same peninsula.

\begin{table*}
    \begin{tabular}{c  c  c  >{\raggedright\arraybackslash}m{120mm}}
            \toprule
src block  &    dst block   &   time        &   traces \\
\hline
c85eb700   &    50f5b000    &   1508630032  & q, 148.245.170.161, 189.209.17.197, 189.209.17.197, 38.104.245.9, 154.24.19.41, 154.54.47.33, 154.54.28.69, 154.54.7.157, 154.54.40.105, 154.54.40.61, 154.54.43.17, 154.54.44.161, 154.54.77.245, 154.54.38.206, 154.54.60.254, 154.54.59.38, 149.6.71.162, 89.228.6.33, 89.228.2.32, 176.221.98.194
\\
\hline
c85eb700   &     50f5b000   &     1508802877 &
q, 148.245.170.161, 200.38.245.45, 148.240.221.29 \\
\bottomrule
    \end{tabular}
    \caption{Traces from the same Ark VPs (mty-mx) to the same destination
            before and during the event block}
    \label{tab:traceroutes}
\end{table*}

We further verify this event by looking at traceroutes.
During the event we see 94 unique Ark VPs attempted 345 traceroutes to the affected blocks.
Of the 94 VPs, 21 VPs (22\%) have their last responsive
  traceroute hop in the same \ac{AS} as the
  target address, and 68 probes (73\%) stopped before reaching that \ac{AS}.
\autoref{tab:traceroutes} shows traceroute data from a single CAIDA Ark VP
  before and during the peninsula described in \autoref{sec:peninsula_definition} and
  \autoref{fig:a30all_raw_50f5b000}.
This data confirms the block was reachable from some locations and not others.
During the event, this trace breaks at the last hop within the source \ac{AS}.

\section{Additional Details about Islands}

\subsection{Country-sized Islands}
	\label{sec:country_sized_islands}

In \autoref{sec:island} we defined islands and gave a sample.
We also have seen country-sized islands.

\begin{figure}
        \includegraphics[width=0.8\columnwidth]{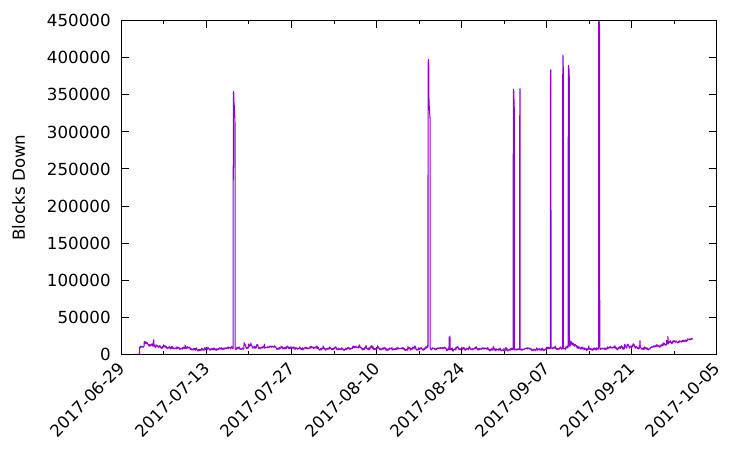}
	\caption{Number of blocks down in the whole responsive Internet. Dataset: A29, 2017q3.}
  \label{fig:a29all_outagedownup_4096}
\end{figure}

In 2017q3 we observed 8 events when it appears that most or all of China
  stopped responding to external pings.
\autoref{fig:a29all_outagedownup_4096} shows
  the number of /24 blocks that were down over time,
  each spike more than 200k /24s,
  between two to eight hours long.
We found no problem reports on network operator mailing lists,
  so we believe these outages were ICMP-specific and likely did not affect
  web traffic.
In addition, we assume the millions of computers inside China continued to operate.
We consider these cases examples of China becoming an \emph{ICMP-island}.

\subsection{Validation of the Sample Island}
\label{sec:additional_island}

In \autoref{sec:island} we reported an island affecting a /24 block where
\ac{VP} E lives.
During the time of the event, E was able to successfully probe addresses within
the same block, however, unable to reach external addresses.
This event started at 2017-06-03t23:06Z, and
can be observed in ~\autoref{fig:islands_plot_down_fraction}.

\begin{figure}
  \includegraphics[width=1\columnwidth]{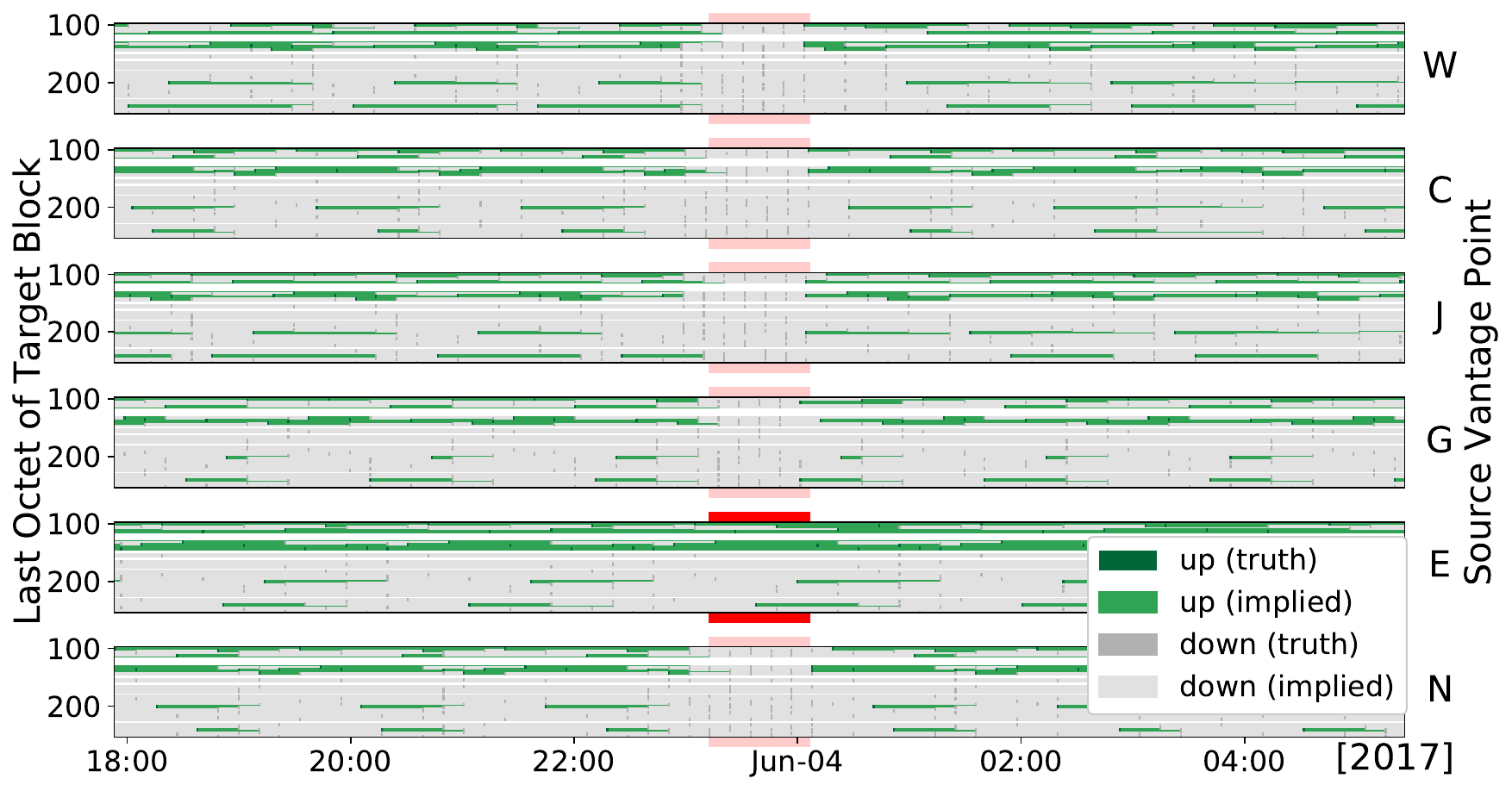}
  \caption{A block
    showing a 1-hour island for this block and \ac{VP} E, while other five
    VPs cannot reach it.}
  \label{fig:a28all_raw_417bca00_6sites}
\end{figure}

Furthermore, no other \ac{VP} was able to reach the affected block for the
time of the island as shown in \autoref{fig:a28all_raw_417bca00_6sites}.

\subsection{Longitudinal View Of Islands}
\label{sec:islands_apendix}

\begin{figure*}
        \begin{center}
          \includegraphics[width=0.85\linewidth]{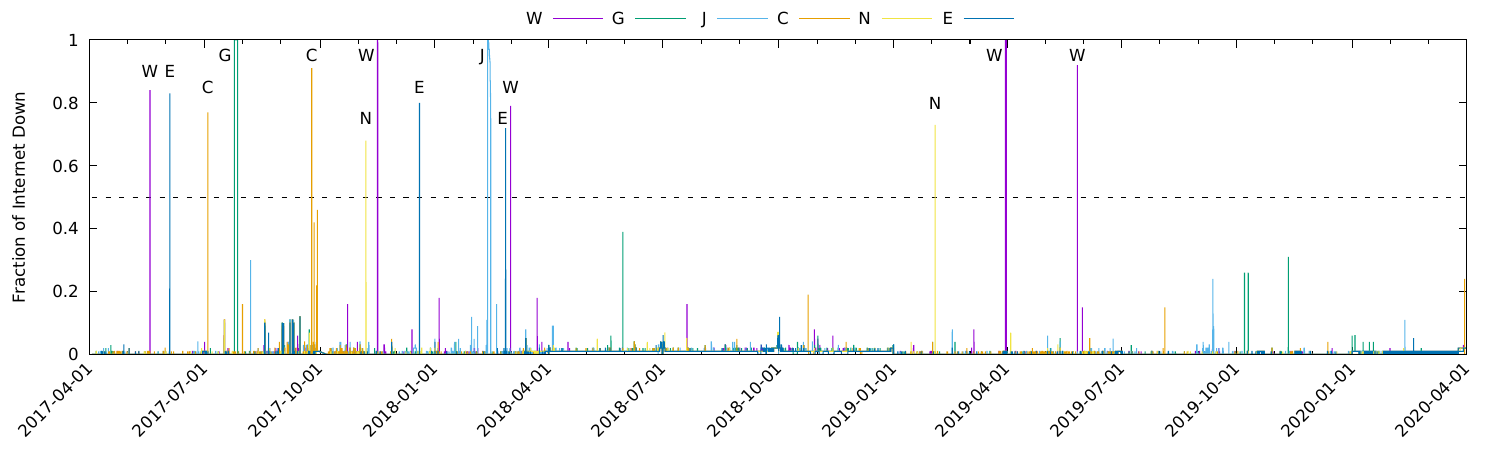}
        \end{center}
        \caption{Islands detected across 3 years using six \acp{VP}.
        Datasets A28-A39.}
  \label{fig:islands_plot_down_fraction}
\end{figure*}

We first consider three years of Trinocular data (described in \autoref{sec:data_sources}),
  from 2017-04-01 to 2020-04-01.
\autoref{fig:islands_plot_down_fraction}
  shows the fraction of the Internet that is
  reachable
  as a dotted line at the 50\% threshold that Chiloe uses to detect an island (\autoref{sec:chiloe}).
We run Chiloe across each VP for this period.

\section{Additional Details about Peninsulas}
\label{sec:additional_peninsula_results}

\subsection{What Sizes Are Peninsulas?}
	\label{sec:peninsula_size}

When network issues cause connectivity problems like peninsulas,
  the \emph{size} of those problems may vary,
  from country-size%
  (see \autoref{sec:country_peninsulas})%
, to \ac{AS}-size,
and also for routable prefixes or fractions of prefixes.
We next examine peninsula sizes.

We begin with Taitao peninsula detection at a /24 block level.
We match peninsulas across blocks within the same prefix by start time and
duration, both measured in one hour timebins.
This match implies that the Trinocular \acp{VP} observing the blocks as up are
also the same.

We compare peninsulas to routable prefixes from Routeviews \cite{routeviews},
  using
  longest prefix matches with /24 blocks.

Routable prefixes consist of many blocks, some of which may not be measurable.
We therefore define the \emph{peninsula-prefix fraction}
  for each routed prefix as fraction of blocks in the peninsula
  that are Trinocular-measurable blocks.
To reduce noise provided by single block peninsulas,
  we only consider peninsulas covering 2 or more blocks in a prefix.

\begin{figure*}
\begin{center}
  \subfloat[Number of Peninsulas]{
    \includegraphics[width=0.65\columnwidth]{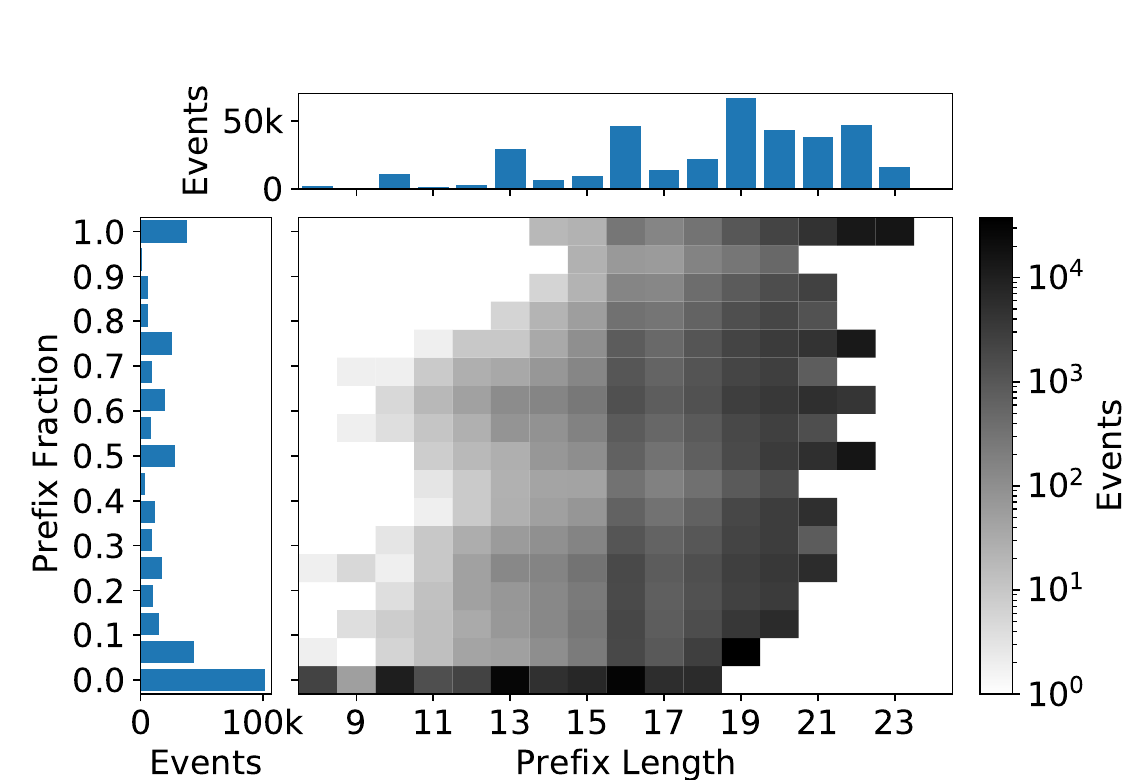}
    \label{fig:a30all_blocks_in_prefix_prefix_fraction_heatmap}
  }
\quad
  \subfloat[Duration fraction]{
    \includegraphics[width=0.65\columnwidth]{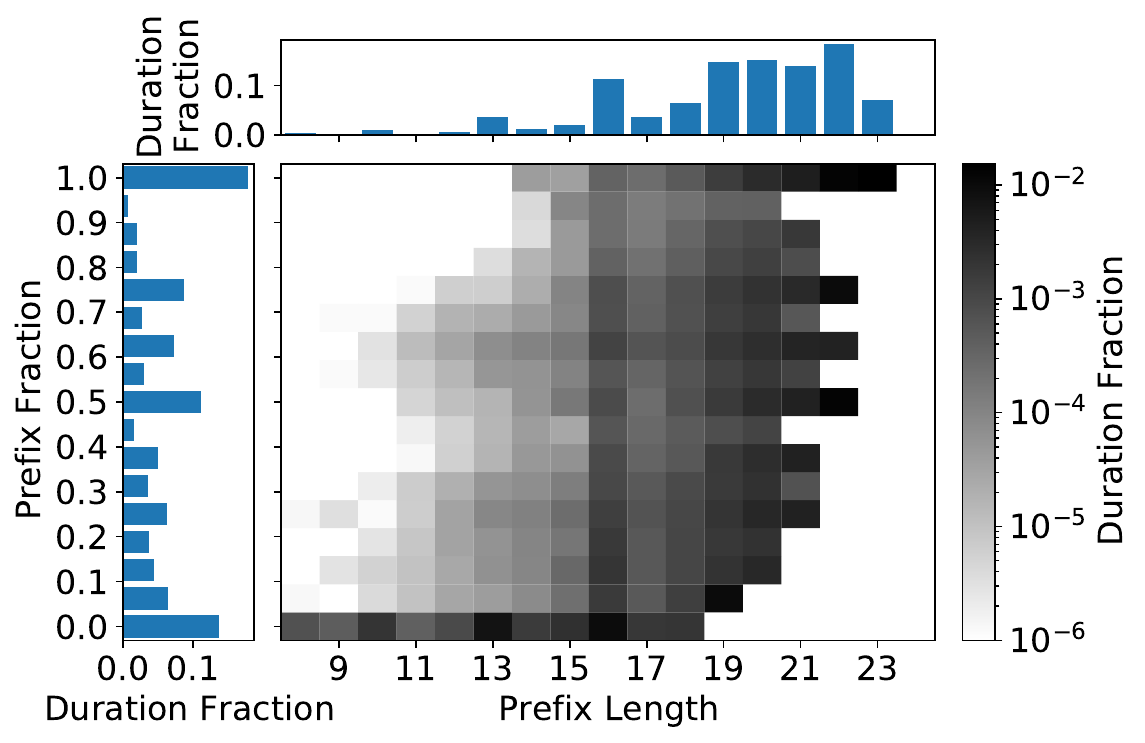}
    \label{fig:a30all_blocks_in_prefix_prefix_fraction_heatmap_duration}
  }
\end{center}
\caption{Peninsulas measured with per-site down events longer than 5 hours. Dataset A30, 2017q4.}
\end{figure*}

\autoref{fig:a30all_blocks_in_prefix_prefix_fraction_heatmap} shows the number
of peninsulas for different prefix lengths and the fraction of the prefix
affected by the peninsula
  as a heat-map,
  where we group them into bins.

We see that about 10\% of peninsulas
  are likely due to
  routing problems or policies,
  since 40k peninsulas affect the whole routable prefix.
However, a third of peninsulas
  (101k, at the bottom of the plot)
  affect
  only a very small fraction of the prefix.
These low prefix-fraction peninsulas suggest
  that they happen \emph{inside} an ISP and
  are not due to interdomain routing.

Finally, we show that \emph{longer-lived peninsulas are likely due to routing or policy choices}.
\autoref{fig:a30all_blocks_in_prefix_prefix_fraction_heatmap_duration}
  shows the same data source,
  but weighted by fraction of time each peninsula
  contributes to the total peninsula time during 2017q4.
Here the larger fraction of weight are peninsulas covering
  full routable prefixes---20\% of all peninsula time during the
quarter (see left margin).

\subsection{Where Do Peninsulas Occur?}
	\label{sec:peninsula_locations}

Firewalls, link failures, and routing problems cause peninsulas on the Internet.
These can either occur inside a given AS,
  or in upstream providers.

To detect where the Internet breaks into peninsulas,
  we look at traceroutes that failed to reach their target address,
  either due to a loop or an ICMP unreachable message.
Then, we find where these traces halt, and
  take note whether halting occurs \emph{at} the target AS and target prefix,
  or \emph{before} the target AS and target prefix.

For our experiment
  we run Taitao to detect peninsulas at target blocks over Trinocular VPs,
  we use Ark's traceroutes~\cite{ark_data_2020} to find last IP address before halt, and
  we get target and halting ASNs and prefixes using RouteViews.

\begin{table}
    \centering
    \footnotesize
    \begin{tabular}{c r r | r r}
      & \multicolumn{2}{c | }{\textbf{Target AS}}
      & \multicolumn{2}{c}{\textbf{Target Prefix}} \\
      Sites Up & At & Before & At & Before \\
      \midrule
      0	& 21,765	    & 32,489	    & 1,775	    & 52,479 \\
      \rowcolor[HTML]{DCDCDC}
      1	& 587	    & 1,197	    & 113	    & 1,671 \\
      \rowcolor[HTML]{DCDCDC}
      2	& 2,981	    & 4,199	    & 316	    & 6,864 \\
      \rowcolor[HTML]{DCDCDC}
      3	& 12,709	    & 11,802	    & 2,454	    & 22,057 \\
      \rowcolor[HTML]{DCDCDC}
      4	& 117,377	& 62,881	    & 31,211	    & 149,047 \\
      \rowcolor[HTML]{DCDCDC}
      5	& 101,516	& 53,649	    & 27,298	    & 127,867 \\
  	\cline{2-5}
      \rowcolor[HTML]{DCDCDC}
      \textbf{1-5} & \cellcolor[HTML]{99ee77} \textbf{235,170} & \textbf{133,728} & \textbf{61,392} & \cellcolor[HTML]{99ee77}\textbf{307,506} \\
      6	& 967,888	& 812,430	& 238,182	& 1,542,136 \\
  \end{tabular}
    \caption{Halt location of failed traceroutes for peninsulas longer than 5
    hours. Dataset A41, 2020q3.}
    \label{tab:peninsula_root_cause}
  \end{table}

In~\autoref{tab:peninsula_root_cause} we show how many traces halt
\emph{at} or \emph{before} the target network.
The center, gray rows show peninsulas (disagreement between \acp{VP})
  with their total sum in bold.
For all peninsulas (the bold row),
  more traceroutes halt at or inside the target AS (235k vs.~134k, the left columns),
  but they more often terminate before reaching the target prefix (308k vs.~61k, the right columns).
This difference suggests policy is implemented at or inside ASes, but not at routable prefixes.
By contrast, outages (agreement with 0 sites up)
  more often terminate before reaching the target AS.
Because peninsulas are more often at or in an AS,
  while outages occur in many places,
  it suggests that peninsulas are policy choices.

\subsection{Country-Level Peninsulas}
	\label{sec:country_peninsulas}

Country-specific filtering is a routing policy made by networks to
restrict traffic they receive.
We next look into  what type of organizations actively block overseas
traffic.
For example, good candidates to restrain who can reach them for security purposes
  are government related organizations.

\begin{table}
    \centering
    \footnotesize
    \begin{tabular}{l c c}
            Industry           & ASes  & Blocks   \\
            \midrule
            ISP               & 23  & 138 \\
            Education         & 21  & 167 \\
            Communications    & 14  & 44  \\
            Healthcare        & 8   & 18  \\
            Government        & 7   & 31  \\
            Datacenter        & 6   & 11  \\
            IT Services       & 6   & 8   \\
            Finance           & 4   & 6   \\
            Other (6 types)
            & \multicolumn{2}{c}{6 (1 per type)}   \\
    \end{tabular}
    \caption{U.S. only blocks. Dataset A30, 2017q4}
    \label{tab:industry}
\end{table}

We test for country-specific filtering (\autoref{sec:detecting_country_peninsulas}) over 2017q4 and find 429
unique U.S.-only blocks in 95 distinct ASes.
We then manually verify each AS categorized by industry
   in \autoref{tab:industry}.
It is surprising how many universities filter by country.
While not common, country specific blocks do occur.

\section{Additional Results}
\label{sec:2020}

{}
Our paper body uses Trinocular measurements for 2017q4 because
  this time period had six active VPs,
  allowing us to make strong statements about how multiple perspectives help.
In this section,
  we verify our results using newer datasets
  to confirm our prior results still hold.
They do---we find quantitatively similar results between 2017 and 2020.

\subsection{Additional Confirmation of the Number of Peninsulas}
\label{sec:additional_confirmation}

\begin{figure*}
\adjustbox{valign=b}{\begin{minipage}[b]{.33\linewidth}
    \includegraphics[width=1\linewidth]{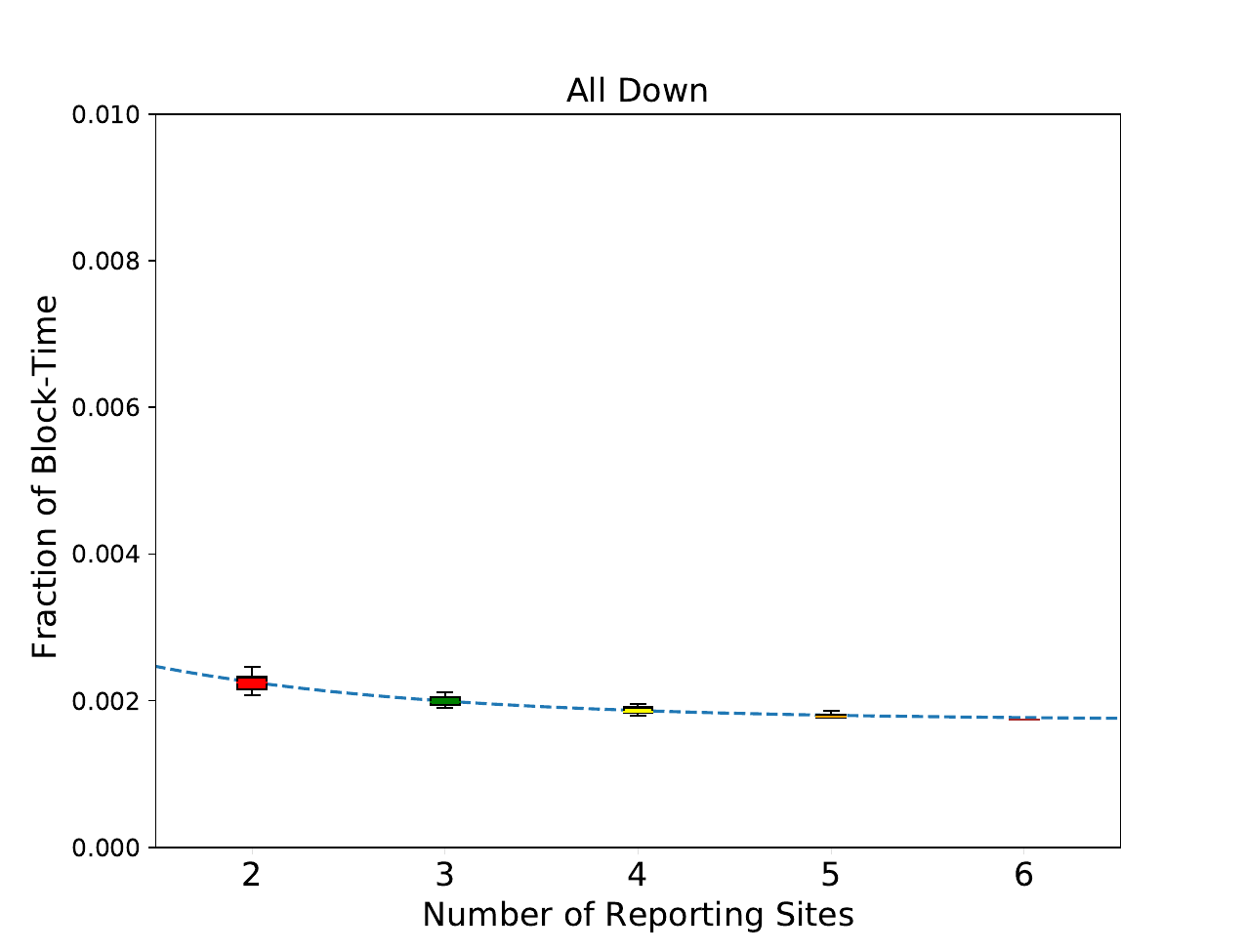}
\end{minipage}}
\adjustbox{valign=b}{\begin{minipage}[b]{.33\linewidth}
    \includegraphics[width=1\linewidth]{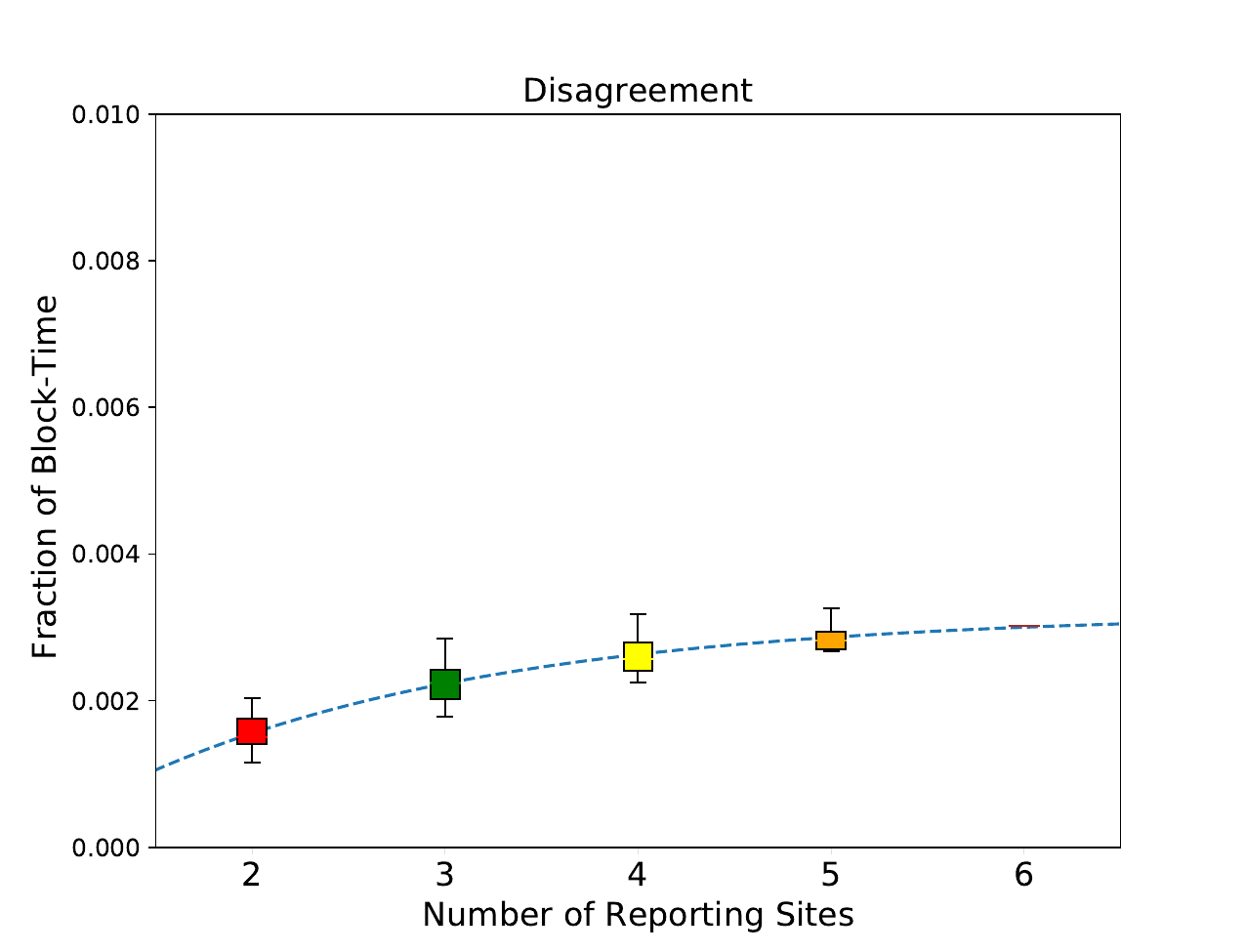}
\end{minipage}}
\adjustbox{valign=b}{\begin{minipage}[b]{.33\linewidth}
    \includegraphics[width=1\linewidth]{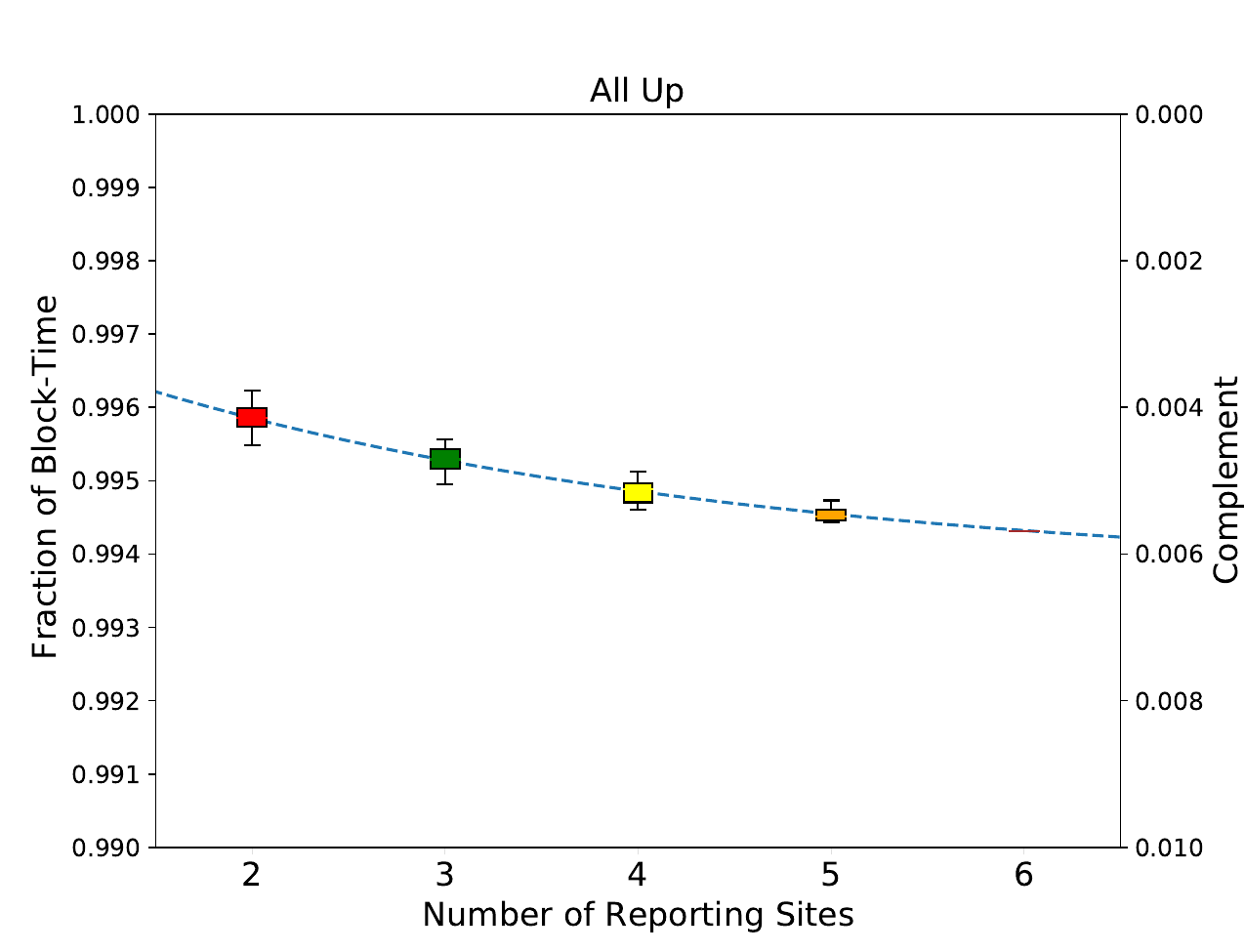}
\end{minipage}}
\caption{Distribution of block-time fraction over sites reporting all down
(left), disagreement (center), and all up (right), for events longer than five
hour. Dataset A34, 2018q4.}
\label{fig:a34all_peninsulas_duration_box}
\end{figure*}

Similarly, as in \autoref{sec:peninsula_frequency},
  we quantify how big the problem of peninsulas is,
  this time using Trinocular 2018q4 data.

In \autoref{fig:a34all_peninsulas_duration_box} we confirm,
  that with more \acp{VP} more peninsulas are discovered,
  providing a better view of the Internet's overall state.

\emph{Outages (left) converge after 3 sites},
  as shown by the fitted curve and decreasing variance.
Peninsulas and all-up converge more slowly.

At six \acp{VP}, here we find and even higher difference between all down and
disagreements.
Confirming that peninsulas are a more pervasive problem than outages.

\subsection{Additional Confirmation of Peninsula Duration}

In \autoref{sec:peninsula_duration} we characterize peninsula duration for
2017q4,
  to determine peninsula root causes.
To confirm our results, we repeat the analysis, but with 2020q3 data.

\begin{figure*}
\begin{center}
  \subfloat [Cumulative events (solid) and duration (dashed)]{
    \includegraphics[width=.63\columnwidth]{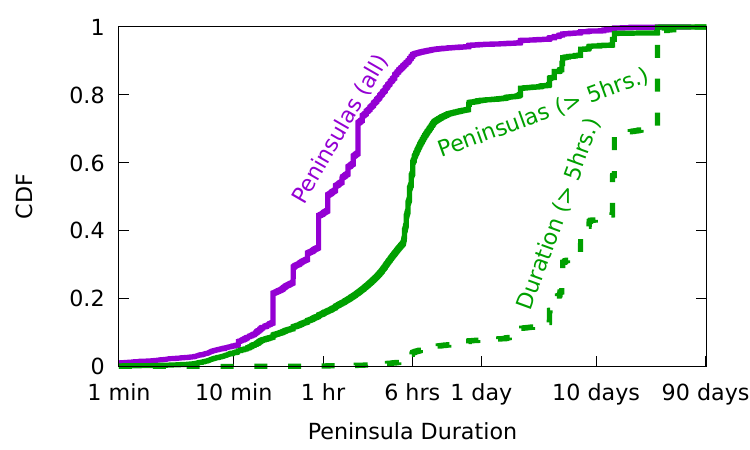}
    \label{fig:a41_partial_outages_duration_cdf}
  }
\quad
  \subfloat[Number of Peninsulas]{
    \includegraphics[width=0.65\columnwidth]{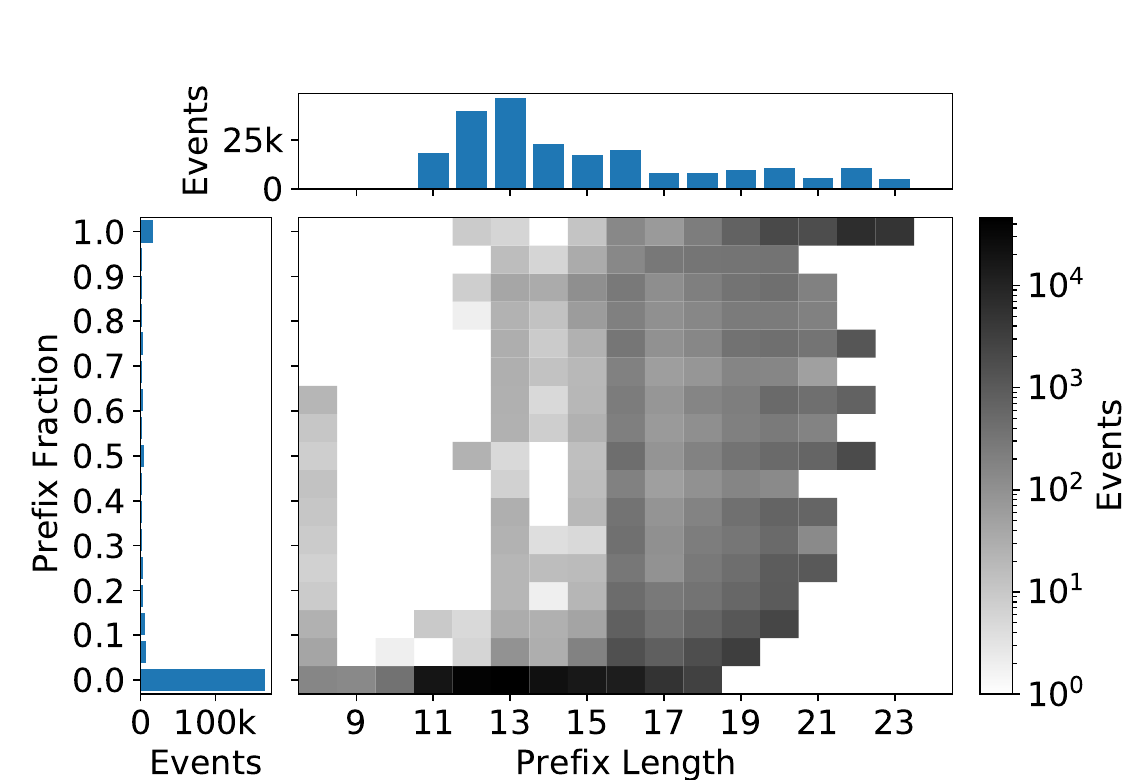}
    \label{fig:a41all_blocks_in_prefix_prefix_fraction_heatmap}
  }
\quad
  \subfloat[Duration fraction]{
    \includegraphics[width=0.65\columnwidth]{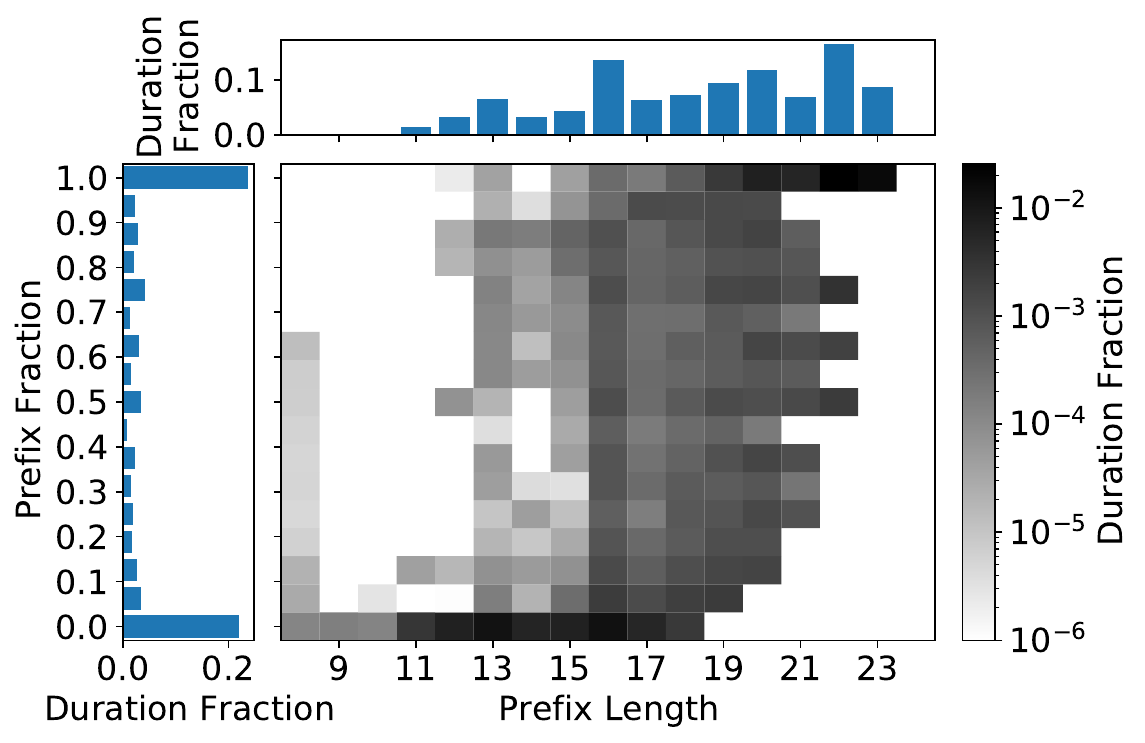}
    \label{fig:a41all_blocks_in_prefix_prefix_fraction_heatmap_duration}
  }
\end{center}
\caption{Peninsulas measured with per-site down events longer than 5 hours during 2020q3. Dataset A41.}
          \label{fig:sim_connection_rollup}
\end{figure*}

As \autoref{fig:a41_partial_outages_duration_cdf} shows,
similarly, as in our 2017q4 results,
  we see that there are many very brief peninsulas (from 20 to 60 minutes).
These results suggest that while the Internet is robust,
there are many small connectivity glitches.

Events shorter than two rounds (22 minutes),
  may represent BGP transients or failures due to random packet loss.

The number of multi-day peninsulas is small,
However, these represent about 90\% of all peninsula-time.
Events lasting a day are long-enough that can be debugged by human network operators,
  and events lasting longer than a week are long-enough that
    they may represent policy disputes.
Together, these long-lived events suggest that
  there is benefit to identifying non-transient peninsulas
  and addressing the underlying routing problem.

\subsection{Additional Confirmation of Size}

In \autoref{sec:peninsula_size} we discussed the size of peninsulas measured as
a fraction of the affected routable prefix.
In the latter section, we use 2017q4 data.
Here we use 2020q3 to confirm our results.

\autoref{fig:a41all_blocks_in_prefix_prefix_fraction_heatmap} shows the peninsulas
per prefix fraction, and \autoref{fig:a41all_blocks_in_prefix_prefix_fraction_heatmap_duration}.
Similarly,
  we find that while small prefix fraction peninsulas are more in numbers,
  most of the peninsula time is spent in peninsulas covering the whole prefix.
This result is consistent with long lived peninsulas being caused by policy
choices.

\label{page:last_page}

\end{document}